\documentclass[final]{siamltex}

\usepackage{epsfig}
\usepackage{amsmath}
\usepackage{amsfonts}

\newtheorem{remarque}{Remark}

\title{The Explicit Simplified Interface Method \\  for compressible
  multicomponent flows.}

\author{Bruno Lombard\thanks{Laboratoire de M\'ecanique et d'Acoustique, 31 chemin Joseph Aiguier, 13402 Marseille, France({\tt lombard@lma.cnrs-mrs.fr}).}
        \and Rosa Donat\thanks{Departament de Matematica Aplicada, Universitat de Valencia,
  46100 Burjassot, Spain({\tt donat@uv.es}).}}

\begin{document}
  
\maketitle

\begin{abstract} 
This paper concerns the numerical approximation of the Euler equations for multicomponent flows. A numerical method is proposed to reduce spurious oscillations that classically occur around material interfaces. It is based on the "Explicit Simplified Interface Method" (ESIM), previously developed in the linear case of acoustics with stationary interfaces (2001, J. Comput. Phys. 168, pp.~227-248). This technique amounts to a higher order extension of the "Ghost Fluid Method" introduced in Euler multicomponent flows (1999, J. Comput. Phys. 152, pp.~457-492). The ESIM is coupled to sophisticated shock-capturing schemes for time-marching, and to level-sets for tracking material interfaces. Jump conditions satisfied by the exact solution and by its spatial derivative are incorporated in numerical schemes, ensuring a subcell resolution of material interfaces inside the meshing. Numerical experiments show the efficiency of the method for rich-structured flows.  
\end{abstract}



\pagestyle{myheadings}
\thispagestyle{plain}
\markboth{B. LOMBARD AND R. DONAT}{INTERFACE METHOD AND EULER FLOWS}

\section{Introduction}

Let us consider multicomponent flows composed of pure inviscid fluids separated by material interfaces. These flows arise in a wide range of physical situations, from water-steam to bubbly flows, liquid suspensions or even high-speed impacts on solids. They may be modeled by the Euler equations, augmented by additional equations describing the fluid composition. The numerical simulation of such configurations leads to major difficulties.

Indeed, even state-of-the-art numerical schemes for single-component flows cannot be applied directly in multicomponent flow simulations. These schemes give rise to oscillations and other computational inaccuracies near material interfaces. For example, any Godunov-type shock-capturing scheme which conserves the mass of individual species fails to maintain pressure equilibrium at a material interface \cite{ABGRALL_96}. The unphysical pressure oscillations stem from the fact that in a shock-capturing scheme, the transition across a material interface is governed by the numerical viscosity of the scheme. When two fluids are involved, intermediate states generated in the numerical transition layer (corresponding to the material interface) are not physically consistent with any component of the mixture: updating the pressure field via one particular equation of state generates erroneous pressure fluctuations. Since material interfaces lack the compressive mechanisms associated to shocks, errors generated within the diffused interface escape and contaminate all flow variables.  

Many numerical methods have been proposed to avoid these unphysical oscillations. See e.g. \cite{ABGRALL_KARNI01} for a concise survey of up-to-date multicomponent methods. Remaining within the {\em front-capturing} (as opposed to {\em front-tracking}) computational framework, we distinguish two main approaches:
 
\underline{\it Miscible models}. Material interfaces are approximated by diffused fronts. An artificial equation of state (or mixture model) is defined, based on thermodynamical arguments \cite{ALLAIRE,LARROUTUROU91,MULET}, which is considered valid for the entire fluid mixture and is used to update the pressure from conserved variables. These models have been analyzed in \cite{ABGRALL_96,ABGRALL_KARNI01,KARNI_94,KARNI_96,SHYUE1}. The elimination of unphysical oscillations generally involves sacrificing strict conservation to some degree, except in \cite{MULET} where a conservative algorithm maintains oscillations under a computationally acceptable level.

\underline{\it Purely inmiscible models}, as in the present paper. A level-set function is used to track material interfaces in an Eulerian manner, i.e. without explicitely computing their location. Depending on the sign of the level-set, one applies the corresponding equation of state. Thermodynamic properties of the fluid change discontinuously across material interfaces, which are preserved as sharp discontinuities. A simple approach was first used in \cite{LSET_EULER}, but it suffers from severe computational inaccuracies in multicomponent flow computations \cite{KARNI_96}.

The {\it Ghost Fluid Method} (GFM) \cite{GFM} is the best-known method belonging to this second group. On each side of a material interface, two fluids are considered: the "real fluid" (i.e. the fluid that really exists on this side) and a "ghost fluid" (i.e. the fluid with the same pressure and velocity than the real fluid on that side, but the entropy of the real fluid on the other side). Near a material interface, classical single-component schemes are then simply applied both on "real-fluid" values and on "ghost-fluid" values. Extensions to multidimensional problems and to other physical situations have been tackled by the GFM \cite{GFM_SOLID,GFM_SHOCK}. However, the GFM suffers from some inaccuracies which are not appreciated when material interfaces separate uniform states. These numerical artifacts are mainly due to the zeroth-order extrapolations used to define the ghost fluid. Simple minded first-order extrapolations do not work \cite{GFM}.

The goal of the present paper is to remove these drawbacks of the GFM by correctly increasing the precision of the computation across a material interface. We propose to implement a carefully designed first-order extrapolation procedure to obtain the "ghost fluid" values. To do so, we adapt the {\it Explicit Simplified Interface Method} (ESIM), previously developed for the linear hyperbolic systems of acoustics with stationary interfaces \cite{LOMBARD1,LOMBARD2,PIRAUX1}. Note that another linear extrapolation in the context of the GFM has been proposed in \cite{MORANO} in order to couple Eulerian and Lagrangian computations. In the present paper, however, the goal is to enforce first-order jump conditions at material interfaces.

Interface methods have been widely used in the numerical treatment of boundaries and interfaces in PDE's: see e.g. the {\it Immersed Interface Method} (IIM), applied to elliptic equations \cite{IIM_LI} and linear hyperbolic systems \cite{IIM_ZHANG}. See \cite{THESE_MOI,PIRAUX1} for a concise survey of up-to-date interface methods. To our knowledge, the present paper is the first attempt to apply interface methods to the nonlinear hyperbolic system of the Euler equations. The computational complexity of the resulting algorithm is essentially the same as that of the GFM, and it can easily be coupled to a wide class of high-order shock-capturing schemes.

The paper is organized as follows. Section 2 recalls the level-set framework to model compressible multicomponent flows. Section 3 describes the numerical schemes. The interface method is detailed in section 4. Numerical tests are proposed in section 5. Conclusions and future works are drawn in section 6.  

\section{The level-set framework for multicomponent flows}

\subsection{The Euler equations}

We focus on multicomponent flows consisting of pure fluids separated by material interfaces. Assuming that all components can be described by a single velocity and pressure function, the flow can be modeled by the compressible Euler equations expressing conservation of mass, momentum, and energy of the fluid mixture. Let $\rho$ be the density of the fluid mixture, $u$ the velocity, $p$ the pressure and $e=\varepsilon+\frac{1}{2}u^2$ the specific total energy, with $\varepsilon$ the specific internal energy. Then, we have
\begin{equation} 
\frac{\textstyle \partial}{\textstyle
  \partial\,t}\,\boldsymbol{U}+\frac{\textstyle \partial}{\textstyle
  \partial\,x}\,\boldsymbol{f(U)}=\boldsymbol{0},  
\label{Euler0} 
\end{equation} 
where the vector of conserved quantities $\boldsymbol{U}$ and the flux function $\boldsymbol{f}$ are   
\begin{equation}
\boldsymbol{U}= 
\left( 
\begin{array}{c} 
\rho\\[5pt]
\rho\,u\\[5pt]
\rho\,e 
\end{array} 
\right),\qquad 
\boldsymbol{f(U)}= 
\left( 
\begin{array}{c} 
\rho\,u\\[5pt]
\rho\,u^2+p\\[5pt]
u\left(\rho\,e+p\right) 
\end{array} 
\right). 
\label{FLUX_EULER} 
\end{equation}
To close the system (\ref{Euler0}), we need to specify the equation of state (EOS). In this paper, we consider the {\it stiffened gas} EOS  
\begin{equation} 
p=(\gamma-1)\,\rho\,\varepsilon-\gamma\,p_{\infty} , 
\label{EOS} 
\end{equation} 
where $p_{\infty}$ is the stiffness parameter. The basic {\it polytropic gas} case is recoverred for $p_{\infty}=0$: in this case, $\gamma$ represents the ratio of specific heats. The EOS (\ref{EOS}) is a reasonable approximation for gases, liquids, and even solids under huge pressure conditions \cite{SAUREL_ABGRALL99}. The sound speed $c$ is given by  
\begin{equation} 
c^2= \frac{\textstyle \gamma\,(p+p_{\infty})}{\textstyle \rho}. 
\label{SSpeed} 
\end{equation} 

\subsection{The level-set equation}

\begin{figure}[htbp] 
\begin{center} 
\includegraphics[scale=0.8]{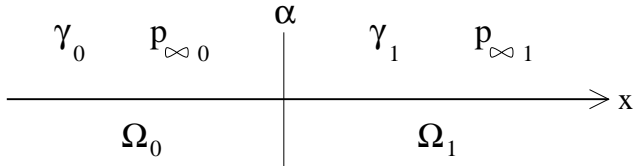} 
\caption{Two components $\Omega_0$ and $\Omega_1$ separated by a material interface.} 
\label{Tampon1} 
\end{center} 
\end{figure}  

The flow description is completed by an additional equation describing the fluid composition. In level-set framework, a scalar function $\phi$ is used to track material interfaces. For the sake of simplicity, we consider only two components $\Omega_0$ and $\Omega_1$, separated by a material interface at $\alpha(t)$ (figure \ref{Tampon1}). We suppose that each fluid satisfies (\ref{EOS}), and that physical parameters of (\ref{EOS}) may be discontinuous at $\alpha$ 
\begin{equation} 
\left(\gamma,p_{\infty}\right)= 
\left\{ 
\begin{array}{l} 
\left(\gamma_0,p_{\infty \, 0}\right) \quad \mbox{ if } x \leq \alpha \\ 
\\ 
\left(\gamma_1,p_{\infty \, 1}\right) \quad \mbox{ if } x > \alpha. 
\end{array} 
\right.
\label{EOS-01}
\end{equation}
The marker variable $\phi$ is initialized as the signed distance function to $\alpha(0)$, which is known initially. Hence its zero level-set defines the material interface at $t=0$, while its sign determines the region occupied by each fluid. 

Since material interfaces propagate with the fluid velocity, the zero level-set of $\phi$ identifies the material interface for all $t>0$ if $\phi(x,t)$ satisfies the advection initial-value problem
\begin{equation} 
\left\{ 
\begin{array}{l} 
\displaystyle 
\frac{\textstyle \partial\,\phi}{\textstyle
  \partial\,t}+u\,\frac{\textstyle \partial\,\phi}{\textstyle
  \partial\,x}=0,
\\  \\ 
\displaystyle 
\phi(x,0)=x-\alpha(0). 
\end{array} 
\right. 
\label{LS:Adv} 
\end{equation}
The evolution equation in (\ref{LS:Adv}) can be recast in conservation form, by combining it with the conservation of mass,
\begin{equation}
\displaystyle 
\frac{\textstyle \partial\, \rho \,\phi}{\textstyle
  \partial\,t}+\frac{\textstyle  \partial\, u\,\rho \,\phi}{\textstyle
  \partial\,x}=0.
\label{LS:Adv.c} 
\end{equation}
Equations (\ref{Euler0}), (\ref{EOS}), and (\ref{LS:Adv}) (or (\ref{LS:Adv.c})) form an "inmiscible" model in which the fluid "mixture" consists of either fluid $\Omega_0$ or $\Omega_1$, and where thermodynamic properties change discontinuously across the material interface.

\section{Numerical schemes}

To integrate the Euler equations and the level-set equation for multicomponent flows, two strategies are available. The first one is to discretize the conservative system composed by the Euler equations (\ref{Euler0}) together with the level-set equation in conservation form (\ref{LS:Adv.c}). This discretization can be performed by applying the classical methodology of shock-capturing schemes for systems of conservation laws. It has been used in other papers (see
e.g. \cite{LSET_EULER}), but the addition of the level-set equation enlarges the Jacobian matrix, which increases the complexity of characteristic decompositions. This feature may become specially cumbersome in multidimensional calculations.

The second strategy - simpler, and followed here - is considered in \cite{GFM}, where the Euler equations (\ref{Euler0}) are discretized independently from the level-set equation in non-conservative form (\ref{LS:Adv}). Following \cite{LSET_ADV}, the Euler system is identified as the {\it minimal system}, hence it can be discretized with any standard conservative scheme, while the level-set equation is independently discretized following its advection form (\ref{LS:Adv}). The numerical flux function involved in the Euler equations will therefore not depend directly on $\phi$, like in a single-component flow. The Lax-Wendroff theorem can still be applied \cite{LSET_ADV}, hence the obtained numerical solution converges to a weak solution of the full conservative system.

The discrete set-up to solve the system (\ref{Euler0})-(\ref{LS:Adv}) is as follows: we consider a lattice of points in the $x$ plane $x_i=i\,\Delta\,x$, where $\Delta\,x$ is a uniform spacing parameter. Numerical values $\boldsymbol{U}_i^n$ and $\phi_i^n$ are respectively considered as approximations to $\boldsymbol{U}$ and $\phi$ at $x_i$ and time $t_n$ \cite{ENO_OSHER}. To decouple the spatial and time integrations, we follow a {\em method of lines} approach \cite{LEV90} for both the Euler system and the level-set equation. The semi-discrete approximation of the nonlinear hyperbolic system (\ref{Euler0}) is written 
\begin{equation} 
\frac{\textstyle d}{\textstyle
  d\,t}\,\boldsymbol{U}_i=\boldsymbol{L}_\Omega(\boldsymbol{U},i),  
\label{SD_U} 
\end{equation} 
whereas the semi-discrete approximation of the advection equation (\ref{LS:Adv}) is
\begin{equation} 
\frac{\textstyle d}{\textstyle
  d\,t}\,\phi_i=G(\phi,i).  
\label{SD_Phi} 
\end{equation} 
The discrete operator $\boldsymbol{L}_\Omega$ in (\ref{SD_U}) is specified in subsection \ref{SEC_EULER}. We say that a grid point is {\it regular} if $\boldsymbol{L}_\Omega$ uses numerical values belonging only to one fluid component. Otherwise, a grid point is called {\it irregular}, and the expression of $\boldsymbol{L}_\Omega$ is modified by the interface method detailed in section 4. The discrete operator $G$ in (\ref{SD_Phi}) is specified in subsection \ref{SEC_LSET}. 

We emphasize the fact that the numerical schemes proposed in the next two subsections are by no means fundamental for the interface method. The reader's favorite single-component solvers can be adapted easily to the forthcoming discussion.    

\subsection{Discretization of the Euler equations}\label{SEC_EULER}

When $x_i$ is a regular point, all the values involved in the flux computations belong to one of the fluid components, say $\Omega_l$. Then, the spatial discrete operator 
$\boldsymbol{L}_\Omega=\boldsymbol{L}_{\Omega_l} $ in (\ref{SD_U}) is written in the customary conservation form 
\begin{equation} 
\boldsymbol{L}_{\Omega_l}(\boldsymbol{U},i)=-\frac{\textstyle
  1}{\textstyle \Delta\,x}  
\left(\boldsymbol{F}_{\Omega_l}
  \left(\boldsymbol{U}_{i-s+1},...,\boldsymbol{U}_{i+s}\right)-  
\boldsymbol{F}_{\Omega_l}\left(\boldsymbol{U}_{i-s},...,\boldsymbol{U}_{i+s-1}\right)\right).     
\label{NumFlu} 
\end{equation} 
The numerical flux function $\boldsymbol{F}_\Omega$ in (\ref{NumFlu}) is the trademark of the scheme; it is defined by a reconstruction procedure, and by a solver. In numerical experiments, we use ENO or WENO reconstructions \cite{ENO_OSHER, WENO}. The width of the stencil $s$ in (\ref{NumFlu}) is related with the theoretical order of accuracy of the spatial reconstruction ($s=3$ for ENO-3 or WENO-5). 

As a solver, we consider a flux-splitting construction \cite{DONAT,PENULTIMATE}. This choice has been considered in \cite{MULET} within the mass fraction model for two-ideal gas flows. This flux-splitting requires two spectral decompositions of the Jacobian at each cell interface, which serve to perform upwind reconstructions of characteristic variables and fluxes. The additional cost is counterbalanced by the robust behavior of the scheme in pathology-prone situations. In addition, and as opposed to a Roe-type numerical flux function, no average-state needs to be computed at a cell interface, which is particularly useful for real-gas simulations. We only need to know the spectral decomposition of the Jacobian matrix: for the EOS (\ref{EOS}), this is given e.g. in \cite{STIRIBA}. 

The time integration of (\ref{SD_U}) is performed by the standard third-order TVD Runge-Kutta \cite{ENO_OSHER}, even when the fifth-order WENO-5 reconstruction is used. $\Delta\,t$ follows from $\Delta\,x$ and from the classical CFL condition of stability  
\begin{equation} 
\mbox{CFL }=\max_{i=0,...,\,N_x} \left(|u_i^n|+c_i^n\right)\,\frac{\textstyle
  \Delta\,t}{\textstyle \Delta\,x}\leq 1, 
\end{equation} 
where $N_x+1$ is the number of grid points.

\subsection{Discretization of the level-set equation}\label{SEC_LSET}

We follow the same method of lines as for the Euler system. The spatial discretization is carried out using the Hamilton-Jacobi framework for the numerical approximation of the derivative terms \cite{HJ_WENO,HJ_ENO}. The upwind direction at $x_i$ is determined by the sign of $u_i$. Full details are in Appendix A.1 of \cite{GFM}.

Theoretically, solving (\ref{LS:Adv}) is sufficient to track the material interface. However, some extra-care must be taken for numerical purposes. Since $u$ is nonuniform along the flow, the numerical representation of $\phi(x,t)$  may become distorted \cite{RUSSO_00}, leading to a poor estimation of the position of $\alpha$. To keep $\phi$ approximately equal to the distance function near the material interface, we follow a classical procedure proposed in \cite{LSET_STOKES}, and called {\it reinitialization} of the level-set. This procedure can be carried out in a number of different ways, here we follow \cite{GFM} and solve to steady state the Hamilton-Jacobi equation  
\begin{equation} 
\left\{ 
\begin{array}{l} 
\displaystyle 
\frac{\textstyle \partial\,{\tilde \phi}}{\textstyle
  \partial\,t}+S({\tilde \phi})\left(\left|\frac{\textstyle
  \partial\,{\tilde \phi}}{\textstyle
  \partial\,x}\right|-1\right)=0,\\  
\\ 
\displaystyle 
{\tilde \phi}(x,0)=\phi(x,t_n), 
\end{array} 
\right. 
\label{LS:Reinit} 
\end{equation}
where $S$ is a smeared sign function given by 
$$ 
S({\tilde \phi})=\frac{\textstyle {\tilde \phi}}{\textstyle
  \sqrt{\textstyle {\tilde \phi}^2+\Delta\,x^2}}.  
$$
The reinitialization equation is solved in fictitious time after each fully complete time step for the Euler equations, with a method of lines approach. For the spatial integration of (\ref{LS:Reinit}), we use a modification of Godunov's method detailed in Appendix A.3 of \cite{GFM}. For the time integration, we use the same TVD Runge-Kutta scheme as for the Euler equations. As a time step, we take $\Delta\,\tau=\Delta\,x$. After five integrations, we obtain a steady distance function, ${\tilde \phi}$, which is then exchanged with $\phi$.  

\section{The ESIM for the Euler equations}

\subsection{The numerical interface treatment}

The ESIM is a numerical treatment to be applied at irregular points, i.e. at points for which the flux computations in the right-hand side of (\ref{SD_U}) involve more that one fluid component. This treatment is carried out at each time integration (such as a Runge-Kutta substep). Let us fix an instant in time (to simplify the notations, the time variable is omitted from now). Then, the basic strategy of the ESIM can be schematically described as follows. 

\begin{figure}[htbp] 
\begin{center} 
\includegraphics[scale=1]{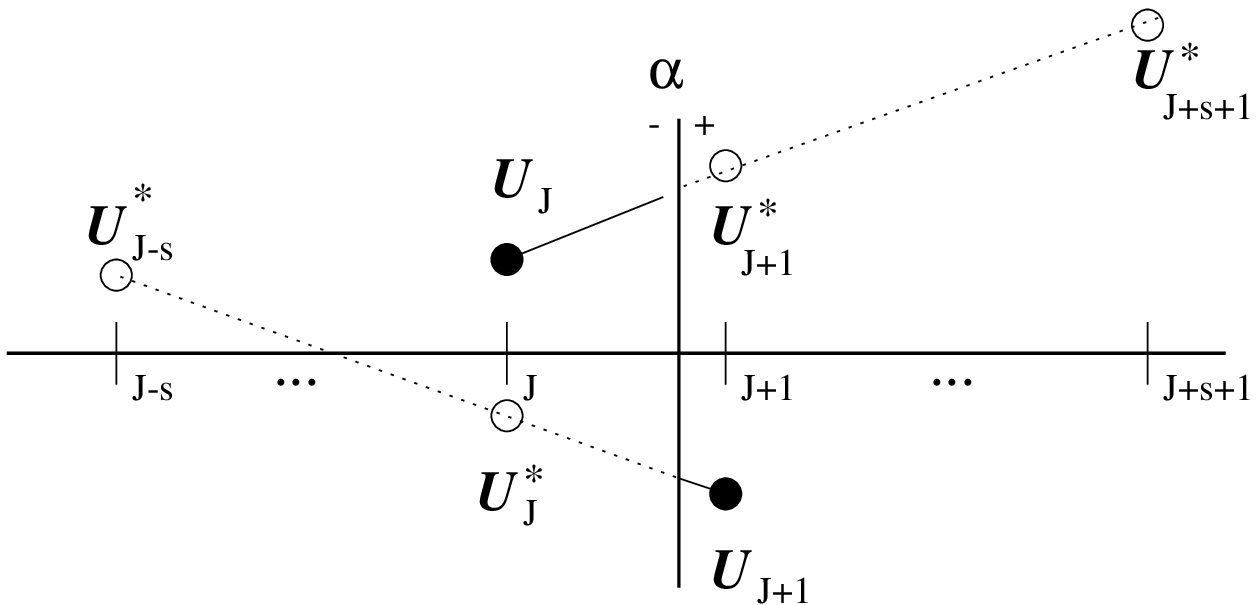} 
\caption{Numerical values $\boldsymbol{U}_i$ and modified values $\boldsymbol{U}^*_i$.} 
\label{Prolonge} 
\end{center} 
\end{figure}  

Consider the material interface at $x=\alpha$ separating the two media $\Omega_0$ and $\Omega_1$ (figure \ref{Tampon1}). We define smooth extensions $\boldsymbol{U}^*(x)$ of the exact solution $\boldsymbol{U}(x)$: here, we consider two-terms Taylor-like expansions past the interface, written 
\begin{equation} 
\begin{array}{l} 
\displaystyle 
\mbox{for } x > \alpha,\qquad
\boldsymbol{U}^*(x)=\boldsymbol{U}(\alpha^-)+(x-\alpha)\,
\frac{\partial}{\partial\,x}\,\boldsymbol{U}(\alpha^-),\\
\\  
\displaystyle 
\mbox{for } x \leq \alpha,\qquad
\boldsymbol{U}^*(x)=\boldsymbol{U}(\alpha^+)+(x-\alpha)\,
\frac{\partial}{\partial\,x}\,\boldsymbol{U}(\alpha^+).   
\end{array}  
\label{SolMod} 
\end{equation}
Then, the interface treatment is divided in two parts.

First, numerical estimations $\boldsymbol{U}_i^*$ of $\boldsymbol{U}^*(x_i)$ in (\ref{SolMod}), called {\it modified values} (or {\it ghost values}, in the GFM framework), are sought at grid points surrounding $\alpha$ (figure \ref{Prolonge}). To do so, one must estimate $\boldsymbol{U}(\alpha^{\pm})$ and $\frac{\partial}{\partial \, x}\,\boldsymbol{U}(\alpha^{\pm})$. These last estimations, based on jump conditions presented in subsection \ref{SEC_JC}, are detailed in subsection \ref{SEC_Upm}.

Second, at each irregular point $x_i$ belonging to a particular fluid component, say $\Omega_l$, the spatial operator $L_{\Omega}(\boldsymbol{U},i)$ is constructed by considering only thermodynamically similar state values, that is: numerical values for grid locations in $\Omega_l$ and modified values at grid points on the other fluid component. This subject is detailed in subsection \ref{SEC_IRR}.
     
\subsection{Jump conditions at material interfaces}\label{SEC_JC}

Across $\alpha$, the primitive variables satisfy the classical zeroth-order jump conditions 
\begin{equation} 
[u]=0,\qquad 
[p]=0,
\label{JC0} 
\end{equation}
where, for any function $f(x,t)$,
$$ 
[f]=\lim_{x \rightarrow \alpha^+}f(x,t)-\lim_{x \rightarrow \alpha^-}f(x,t).
$$
Since $u$ and $p$ do not jump across the interface, then their material - or Lagrangian - derivatives do not either. This fact immediately leads to the less-classical first-order jump conditions
\begin{equation} 
\left[\frac{\textstyle 1}{\textstyle \rho}\,\frac{\textstyle
    \partial\,p}{\textstyle \partial\,x}\right]=0, \qquad
\left[\rho\,c^2\,\frac{\textstyle \partial\,u}{\textstyle
    \partial\,x}\right]=0.  
\label{JC1} 
\end{equation} 
One can also find second-order (and higher) jump conditions. However and unlike the acoustics case \cite{PIRAUX1,THESE_MOI}, these jump conditions (not presented here) involve nonlinear combinations of $p$, $u$ and their successive spatial derivatives, making the computations much more intricate.
 
For polytropic gases, $\rho \,c^2 = \gamma \,p$; since  $[p]=0$, the last equation of (\ref{JC1}) can be simplified in $[\gamma\, \frac{\partial\,u}{\partial\,x}]=0$. Note also that nothing is said about $[\rho]$ and $[\frac{\partial\,\rho}{\partial\,x}]$. 

\subsection{Numerical estimation of one-sided quantities}\label{SEC_Upm}

In a level-set model, the location of the material interface is given by a sign change in the marker variable $\phi$. Suppose that 
\begin{equation} 
\phi_J \times \phi_{J+1} <0.
\label{PhiTest} 
\end{equation}
As a consequence, the interface lies somewhere between $x_J$ and $x_{J+1}$. The subcell location of $\alpha$ is specified by the parameter
\begin{equation} 
\tilde \theta=\frac{\textstyle\alpha-x_J}{\textstyle \Delta\,x} \, \in [0,1[,
\label{Theta} 
\end{equation} 
which is readily estimated to second-order by a simple linear interpolation involving the smooth level-set function $\phi$. Indeed, defining
\begin{equation} 
\theta=\frac{\textstyle |\phi_J|}{\textstyle |\phi_J|+|\phi_{J+1}|},
\label{Theta-Phi} 
\end{equation} 
we have $\theta= \tilde \theta + O(\Delta x^2)$. Since the jump conditions (\ref{JC0}) and (\ref{JC1}) concern the primitive variables $\boldsymbol{W}=\,^T\left(\rho,\,u,\,p\right)$, it is easier to first look for estimations of $\boldsymbol{W}(\alpha^{\pm})$ and $\frac{\partial}{\partial \,x}\,\boldsymbol{W}(\alpha^{\pm})$, and then to go back to the conserved variables $\boldsymbol{U}(\alpha^{\pm})$ and $\frac{\partial}{\partial\,x}\,\boldsymbol{U}(\alpha^{\pm})$. In the following discussion and for any function $f$, the numerical estimation of $f(\alpha^{\pm})$ or $\frac{\partial}{\partial\,x}\,f(\alpha ^{\pm})$ is denoted by
$f^{\pm}$ or $f_x^{\pm}$. 
  
{\it Density}. The density is not subject to any constraint at a material interface, therefore numerical estimations $\rho^{\pm}$ and $\rho^{\pm}_x$ are performed via one-sided interpolations. On the left of $\alpha$, elementary interpolations lead to 
\begin{equation} 
\begin{array}{l} 
\displaystyle 
\rho^-=-\theta\,\rho_{J-1}+(1+\theta)\,\rho_J,\\[10pt]  
\displaystyle 
\rho^-_x =\frac{\textstyle 1}{\textstyle \Delta\,x}\left(\rho_J-\rho_{J-1}\right), 
\end{array} 
\label{RhoM} 
\end{equation} 
while on the right of $\alpha$, we get
\begin{equation} 
\begin{array}{l} 
\displaystyle 
\rho^+=(2-\theta)\,
\rho_{J+1}-(1-\theta)\,\rho_{J+2},\\[10pt]
\displaystyle 
\rho^+_x=\frac{\textstyle 1}{\textstyle \Delta\,x}\,\left(\rho_{J+2}- \rho_{J+1}\right).  
\end{array} 
\label{RhoP} 
\end{equation} 
If $\rho(x)$ is a piecewise $C^1$ function, we obviously have
\[ \rho(\alpha^{\pm})= \rho^{\pm} + O(\Delta\,x^2), \qquad
\frac{\textstyle \partial\, \rho}{\textstyle  \partial\,x}(\alpha^{\pm})=\rho^{\pm}_x + O(\Delta\,x). \]

\begin{remarque} The density estimations obtained above must remain within the physical range, i.e. $\rho^{\pm}>0$. In some cases, negative values can be numerically obtained (e.g. in regions close to vacuum). Then, a different approximation procedure with constraints would be necessary (which is not considered here).
\end{remarque}

{\it Pressure}. The pressure and its spatial derivative satisfy (\ref{JC0}) and (\ref{JC1})  
\begin{equation} 
\begin{array}{c} 
\displaystyle 
p(\alpha^+)=p(\alpha^-),\\[10pt]
\displaystyle 
\frac{\textstyle 1}{\textstyle \rho(\alpha^+)}\,\frac{\textstyle
  \partial\,p}{\textstyle \partial\,x}(\alpha^+)=\frac{\textstyle
  1}{\textstyle \rho(\alpha^-)}\,\frac{\textstyle
  \partial\,p}{\textstyle \partial\,x}(\alpha^-).  
\end{array} 
\label{JCp} 
\end{equation} 
From the definition of $p^{\pm}$ and $p_x^{\pm}$, we have
\begin{equation}  \label{sysp1}
\begin{array}{lll} 
p_J&=& \displaystyle p^-+(x_J-\alpha)\,p_x^-,\\[10pt]
p_{J+1}&=& \displaystyle p^+ +(x_{J+1}-\alpha)\,p_x^+.  
\end{array} 
\end{equation}
We impose that numerical estimations satisfy exactly the same jump conditions as exact values. Combining (\ref{sysp1}) with (\ref{JCp}) leads to the $2
\times 2$ system  
\begin{equation} 
\left( 
\begin{array}{c} 
p_J\\[10pt]
p_{J+1} 
\end{array} 
\right) 
= 
\left( 
\begin{array}{cc} 
1 & -\theta\,\Delta\,x\\[10pt]
1 & \displaystyle (1-\theta)\,\frac{\textstyle \rho^+}{\textstyle
 \rho^-}\,\Delta\,x   
\end{array} 
\right) 
\left( 
\begin{array}{c} 
p^-\\[10pt]
\displaystyle 
p_x^- 
\end{array}  
\right). 
\end{equation} 
The above matrix is always invertible, provided that $\rho^{\pm}>0$. Inverting the above system gives the numerical values at $\alpha^-$ 
\begin{equation} 
\begin{array}{l} 
\displaystyle 
p^-=\frac{\textstyle 1}{\textstyle \displaystyle
  (1-\theta)\,\frac{\textstyle \rho^+}{\textstyle
    \rho^-} + \theta}\,\left((1-\theta)\,\frac{\textstyle
  \rho^+}{\textstyle \rho^-}\,p_J +
\theta\,p_{J+1}\right),\\[30pt]
\displaystyle 
p_x^- =\frac{\textstyle 1}{\textstyle \displaystyle
  \left((1-\theta)\,\frac{\textstyle \rho^+}{\textstyle
    \rho^-} +
  \theta\right)\,\Delta\,x}\,\left(p_{J+1}-p_J\right).
\end{array} 
\label{PreM} 
\end{equation} 
To get estimations at $\alpha^+$, one uses (\ref{PreM}) and jump conditions (\ref{JCp}). Note that $p^+$ (or $p^-$) is a convex combination with positive coefficents of $p_J$ and $p_{J+1}$, hence $p^{\pm}>0$. As in the previous case, standard results allow to conclude that if both $p(x)$ and $\rho(x)$ are $C^1$ away from the material interface, then
\[ p(\alpha^\pm)= p^{\pm} +  O(\Delta\,x^2), \qquad
\frac{\partial p}{\partial x}(\alpha^{\pm}) = p_x^{\pm} + O(\Delta\,x).\]

{\it Velocity}. Zeroth-order and first-order jump conditions for the velocity are 
\begin{equation} 
\begin{array}{c} 
\displaystyle 
u(\alpha^+)=u(\alpha^-),\\[10pt]
\displaystyle 
\rho(\alpha^+)\,c^2(\alpha^+)\,\frac{\textstyle
  \partial\,u}{\textstyle
  \partial\,x}(\alpha^+)=\rho(\alpha^-)\,c^2(\alpha^-)\,\frac{\textstyle
  \partial\,u}{\textstyle \partial\,x}(\alpha^-).  
\end{array} 
\label{JCu} 
\end{equation}
Applying the same procedure as for the pressure, we obtain a system which is invertible  as long as $ (1-\theta)\gamma_0\left(p^-+p_{\infty\,0}\right) / \gamma_1\left(p^++p_{\infty \, 1}\right) +\theta >0$. This is always the case since $p^{\pm}>0$. For stiffened gas EOS (\ref{EOS}), defining 
\begin{equation} 
\xi=\frac{\textstyle  \gamma_0\left(p^-+p_{\infty \,0}\right)}
  {\textstyle \gamma_1\left(p^-+p_{\infty \, 1}\right)},
\label{XI} 
\end{equation}
(for the ideal polytropic case $\xi=\gamma_0/\gamma_1$) with $p^-$ given by (\ref{PreM}), we obtain
\begin{equation} 
\begin{array}{l} 
\displaystyle 
u^-=\frac{\textstyle 1}{\textstyle \displaystyle
  (1-\theta)\,\xi + \theta}\,\left((1-\theta)\,\xi\,u_J +
\theta\,u_{J+1}\right), \\[20pt]
\displaystyle 
u_x^-=\frac{\textstyle 1}{\textstyle \displaystyle
  \left((1-\theta)\,\xi +
  \theta\right)\,\Delta\,x}\,\left(u_{J+1}-u_J\right). 
\end{array} 
\label{VitM} 
\end{equation} 
To get estimations at $\alpha^+$, one uses (\ref{VitM}) and the jump conditions (\ref{JCu}). As before, under the appropriate smoothness assumptions, one has 
\begin{equation}
u(\alpha^{\pm}) = u^{\pm} +O(\Delta\,x^2), \qquad
\frac{\partial u}{\partial x}(\alpha^{\pm})= u_x^\pm + O(\Delta\,x).
\end{equation}

{\it Conserved variables}.
Once sided estimations $\boldsymbol{W}^{\pm}$ and $\boldsymbol{W}_x^{\pm}$ are computed, we deduce $\boldsymbol{U}^{\pm}$ and $\boldsymbol{U}_x^{\pm}$ from (\ref{EOS}). These values are second-order approximations to the exact values under appropriate smoothness assumptions, hence
\begin{equation} \label{Ustar}
\boldsymbol{{\hat U}}(x)= 
\left \{ 
\begin{array}{ll}
 \boldsymbol{U}^+ + (x-\alpha)\, \boldsymbol{U}_x^+ & \mbox{ for } x\leq \alpha \\
\\
 \boldsymbol{U}^- + (x-\alpha)\, \boldsymbol{U}_x^- & \mbox{ for } x>
 \alpha
 \end{array} \right .
\end{equation}
satisfies $\boldsymbol{U}^*(x)=\boldsymbol{{\hat U}}(x)+ O(\Delta x^2)$ near the interface. 

\subsection{Spatial integration at irregular points}\label{SEC_IRR}

From (\ref{NumFlu}) and (\ref{PhiTest}), one deduces that the irregular points are 
\begin{equation}
x_{J-s+1},...,\,x_{J+s}. 
\label{Irregular}
\end{equation}
Modified values must also be computed at $x_{J-s}$ and $x_{J+s+1}$ because the material interface can move between $t_n$ and $t_{n+1}$, crossing one grid point (but no more, because of the CFL condition): as a consequence, these two regular points can become irregular points during one Runge-Kutta substep. The modified values are computed by substituting the appropriate grid point in the expresion (\ref{Ustar}). Using (\ref{Theta-Phi}) as a second-order approximation to (\ref{Theta}),
we obtain   
\begin{equation} 
\begin{array}{lll} 
i=J+1,...,J+s+1,\qquad &\boldsymbol{U}^*_i&=\displaystyle \boldsymbol{U}^-+\left(i-J-\theta\right)\,\Delta\,x\,\boldsymbol{U}^-_x,\\ 
\\ 
i=J-s,...,J,\qquad &\boldsymbol{U}^*_i&=\displaystyle \boldsymbol{U}^++\left(i-J-\theta\right)\,\Delta\,x\,\boldsymbol{U}^+_x.  
\end{array} 
\label{UI*} 
\end{equation} 
At each irregular point $x_i$, $\boldsymbol{L}_\Omega$ is now applied on numerical values on the same side than $x_i$, and on modified values on the other side than $x_i$, hence
\begin{equation}
\begin{array}{l}
i=J-s+1,...,\,J,\\ 
\\
\displaystyle
\quad \boldsymbol{L}_\Omega(\boldsymbol{U},i)=-\frac{\textstyle
  1}{\textstyle \Delta\,x}  
\left(\boldsymbol{F}_{\Omega\,0}\left(\boldsymbol{U}_{i-s+1},...,\boldsymbol{U}_{J}, \boldsymbol{U}_{J+1}^*,...,\boldsymbol{U}_{i+s}^*\right) \right.\\ 
\\ \qquad \qquad \qquad \qquad
\displaystyle \left.
- \boldsymbol{F}_{\Omega \,0}\left(\boldsymbol{U}_{i-s},...,\boldsymbol{U}_{J},\boldsymbol{U}_{J+1}^*,...,\boldsymbol{U}_{i+s-1}^* \right)\right), \\
\\  
i=J,...,\,J+s,\\ 
\\
\displaystyle
\quad \boldsymbol{L}_\Omega(\boldsymbol{U},i)=-\frac{\textstyle
  1}{\textstyle \Delta\,x}  
\left(\boldsymbol{F}_{\Omega\,1}\left(\boldsymbol{U}_{i-s+1}^*,...,\boldsymbol{U}_{J}^*, \boldsymbol{U}_{J+1},...,\boldsymbol{U}_{i+s}\right) \right. \\
\\ \qquad \qquad \qquad \qquad
\displaystyle \left.
- \boldsymbol{F}_{\Omega \,1}\left(\boldsymbol{U}_{i-s}^*,...,\boldsymbol{U}_{J}^*,\boldsymbol{U}_{J+1},...,\boldsymbol{U}_{i+s-1} \right)\right).   
\end{array}
\label{TM_ESIM}
\end{equation}

\subsection{Summary of the implementation} \label{ALGO} 

Suppose that Runge-Kutta integrations (denoted by RK) have been performed up to the $(m-1)$-th substep. Then, the time-marching for both the Euler equations and the level-set equation can be summed up in 8 steps. 

{\bf Step 1}: {\it location of the interface}. Compute $J$ (\ref{PhiTest}) and $\theta$ (\ref{Theta}). 

{\bf Step 2}: {\it construction of modified values}. Compute (\ref{UI*}).  

{\bf Step 3}: {\it construction of temporary values}. Build two sets $\boldsymbol{A}_i$ and $\boldsymbol{B}_i$
\begin{equation} 
\boldsymbol{A}_i^{(m-1)}= 
\left\{ 
\begin{array}{l} 
\boldsymbol{U}_i^{(m-1)},\, i=0,...,J,\\ 
\\ 
\boldsymbol{U}_i^*,\, i=J+1,...,J+s+1, 
\end{array} 
\right.\\ 
\boldsymbol{B}_i^{(m-1)}= 
\left\{ 
\begin{array}{l} 
\boldsymbol{U}_i^*,\, i=J-s,...,J,\\ 
\\ 
\boldsymbol{U}_i^{(m-1)},\, i=J+1,...,N_x. 
\end{array} 
\right. 
\label{AiBi} 
\end{equation} 

{\bf Step 4}: {\it update $\boldsymbol{A}_i$ and $\boldsymbol{B}_i$}. Compute one Runge-Kutta substep of (\ref{SD_U})
\begin{equation} 
\begin{array}{lll} 
i=s,...,J+1,\qquad  
&\boldsymbol{A}_i^{(m)}&=\mbox{ RK }\left(\boldsymbol{A}_i^{(k)}, \boldsymbol{L}_{\Omega_0}\right),\quad k \leq m\\ 
&&\\ 
i=J,...,N_x-s, \qquad  
&\boldsymbol{B}_i^{(m)}&=\mbox{ RK }\left(\boldsymbol{B}_i^{(k)}, \boldsymbol{L}_{\Omega_1}\right),\quad k \leq m. 
\end{array} 
\label{RK_ESIM} 
\end{equation}  

{\bf Step 5}: {\it update $\phi$.} Compute one Runge-Kutta substep of (\ref{LS:Adv}) via $u^{(m-1)}$. 
 
{\bf Step 6}: {\it location of the interface.} Compute $J$ (\ref{PhiTest}) from $\phi^{(m)}$. 
 
{\bf Step 7}: {\it update $\boldsymbol{U}$.} 
From (\ref{RK_ESIM}) and $J$, select 
\begin{equation} 
\boldsymbol{U}_i^{(m)}= 
\left\{ 
\begin{array}{l} 
\boldsymbol{A}_i^{(m)} \mbox{ for } i=0,...,J,\\ 
\\ 
\boldsymbol{B}_i^{(m)} \mbox{ for } i=J+1,...,N_x. 
\end{array} 
\right. 
\end{equation} 

{\bf Step 8}: {\it reinitialization of $\phi$.} 
If $m=3$ (i.e. $t=t_{n+1}$), integrate (\ref{LS:Reinit}). 

\subsection{Some remarks}\label{SEC_REMARK} 
 
{\it Complexity}. The algorithm presented in section \ref{ALGO} is simple. The computation of modified values $\boldsymbol{U}_i^*$ does not depend on the discrete spatial operator. No analytical results - such as the solution of a Riemann problem - are required. As deduced from (\ref{RK_ESIM}), one does not need to write a new solver: the adaptation of known single-component solvers to the multicomponent case is direct. The only difficulty is to switch precisely modified values in the appropriate solver, depending on the sign of the level-set function, which is in fact an easy task. 
 
{\it Computational cost}. In comparison with single-component simulations, our approach leads to a +25 \% additional cost, both on a memory and computational time point of view. This cost is almost completely due to the level-set function $\phi$. Such a cost is inherent to level-set formulations, in order to know the composition of the fluid at each grid point. Eulerian methods that do not use level-sets require in counterpart to modify the Euler eigenstructure, leading to a similar additional cost.    
  
{\it Consistency}. As for the GFM \cite{ABGRALL_KARNI01}, the ESIM works because values used for time-stepping in (\ref{RK_ESIM}) are thermodynamically similar: $\boldsymbol{A}_i^*$ (respectively $\boldsymbol{B}_i^*$) in (\ref{AiBi}) satisfy the same equation of state. The algorithm amounts to consider separately two single-component flows, where no oscillations exist.  
 
{\it Single-component flow}. In the limit case $\gamma_0=\gamma_1$, $p_{\infty\,0}=p_{\infty\,1}$, the flow is single-component, and material interfaces amount to classical contact-discontinuities. The ESIM behaves equally well, sharpening contact discontinuities.  
 
{\it Conservativity}. The ESIM is formally non-conservative locally, since
\begin{equation} 
\boldsymbol{F}_{\Omega_0}\left(\boldsymbol{U}_{J-s+1},...,\boldsymbol{U}_J,
\boldsymbol{U}_{J+1}^*,...,\boldsymbol{U}_{J+s}^*\right)\neq   
\boldsymbol{F}_{\Omega_1}\left(\boldsymbol{U}_{J-s+1}^*,...,\boldsymbol{U}_J^*, \boldsymbol{U}_{J+1},...,\boldsymbol{U}_{J+s}\right),\\ 
\label{NonCons}  
\end{equation} 
like in \cite{KARNI_94} or \cite{GFM}. However, the lack of conservation is only introduced at one cell boundary on the entire domain. In practice, this feature does not seem to spoil convergence to the correct solution. 

Note that a fully-conservative version of the GFM has been developed \cite{GFM_CONS}. Such an approach could probably be adapted to the ESIM, but we do not look further in the present paper: unlike in \cite{GFM_CONS} where inert shocks and detonation waves are adressed, we only focus on the material interfaces (where conservation errors are not crucial).

\section{Numerical experiments} \label{numex}
 
\subsection{Configurations} 
 
Four numerical experiments are proposed. Tests 1 and 2 illustrate the interaction of shock waves with one and two material interfaces. Test 3 is a pure advection problem, with smooth structures. Test 4 concerns nonlinear acoustics. Analytical values and numerical values are respectively shown in solid lines and dotted lines, and we take CFL=0.66. 
 
All tests have initially-isolated material interfaces. Riemann problems are not our concern, since jump conditions (\ref{JC0}) - which are a building-block of the ESIM - are then generally not satisfied. Note that the GFM has the same limitation. In practice, the GFM often works for Riemann problems, but some failures (see e.g. data of Test 4 in \cite{ABGRALL_KARNI01}) can be explained by the fact that basic assumptions are not satisfied. 
   
\subsection{Test 1: shock-interface interaction}\label{Sec_Num1}
 
\begin{figure}[htbp] 
\begin{center} 
\begin{tabular}{cc} 
Density & Velocity\\ 
\includegraphics[width=5.5cm,height=5.5cm]{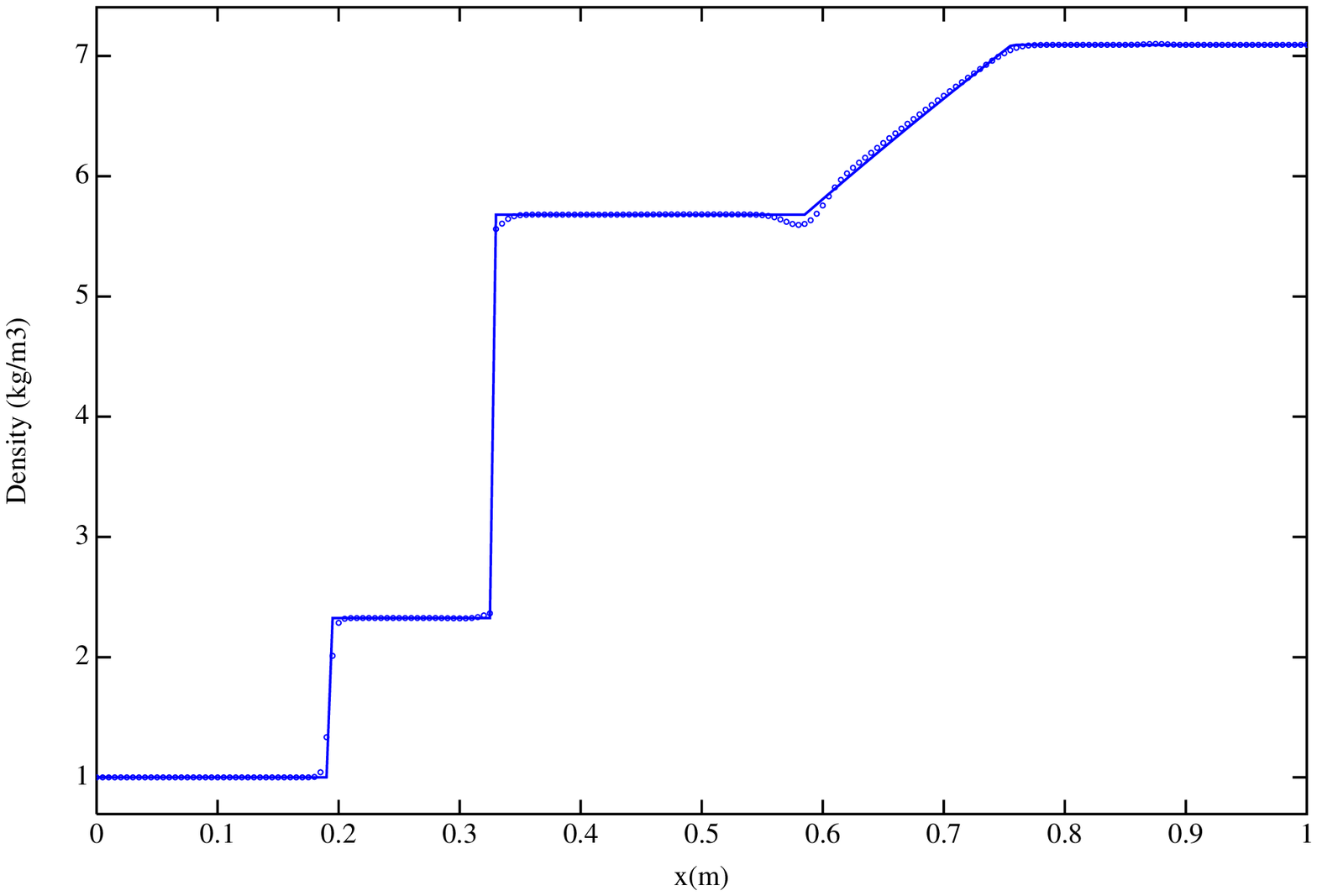}&  
\includegraphics[width=5.5cm,height=5.5cm]{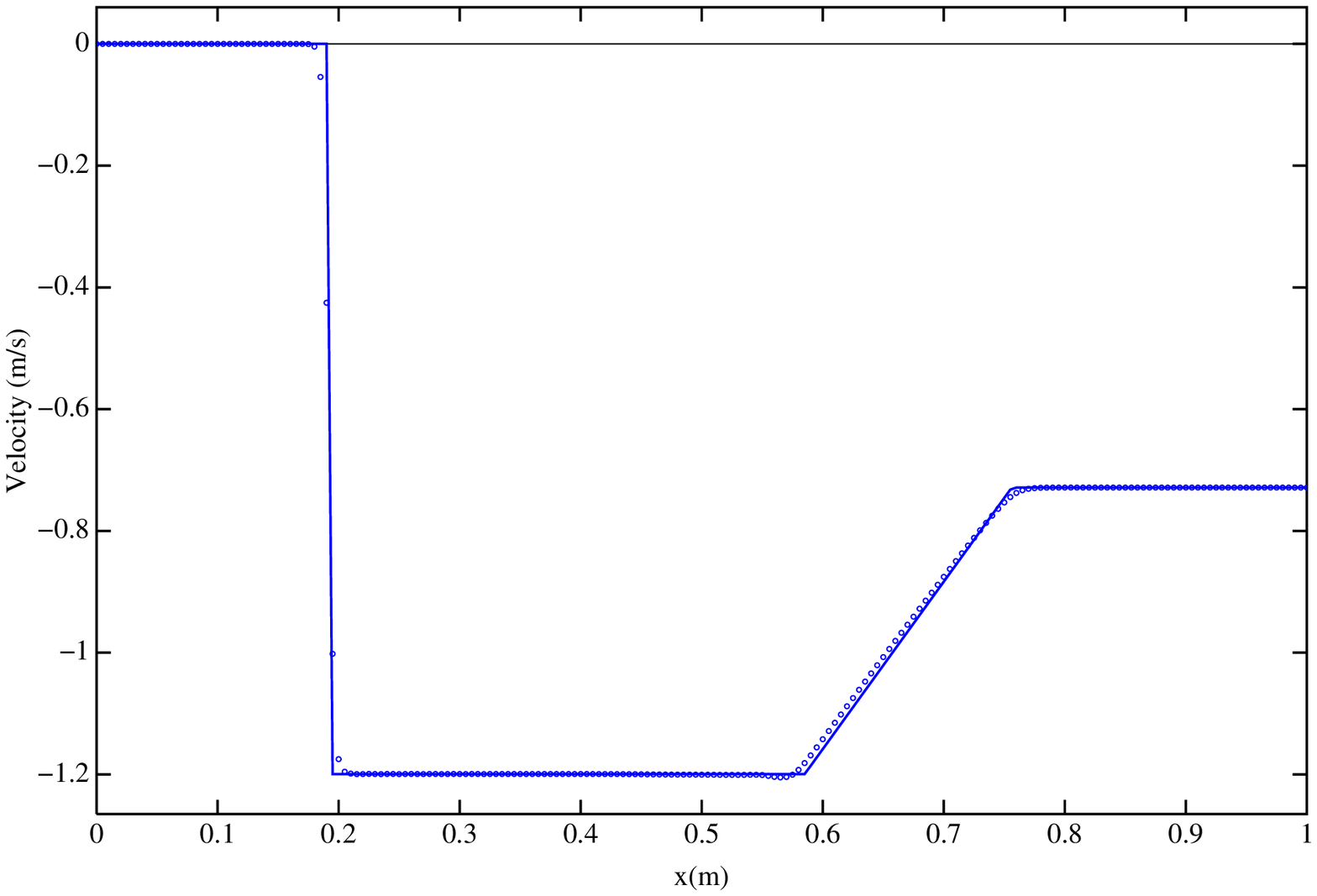}\\ 
Entropy & Pressure\\ 
\includegraphics[width=5.5cm,height=5.5cm]{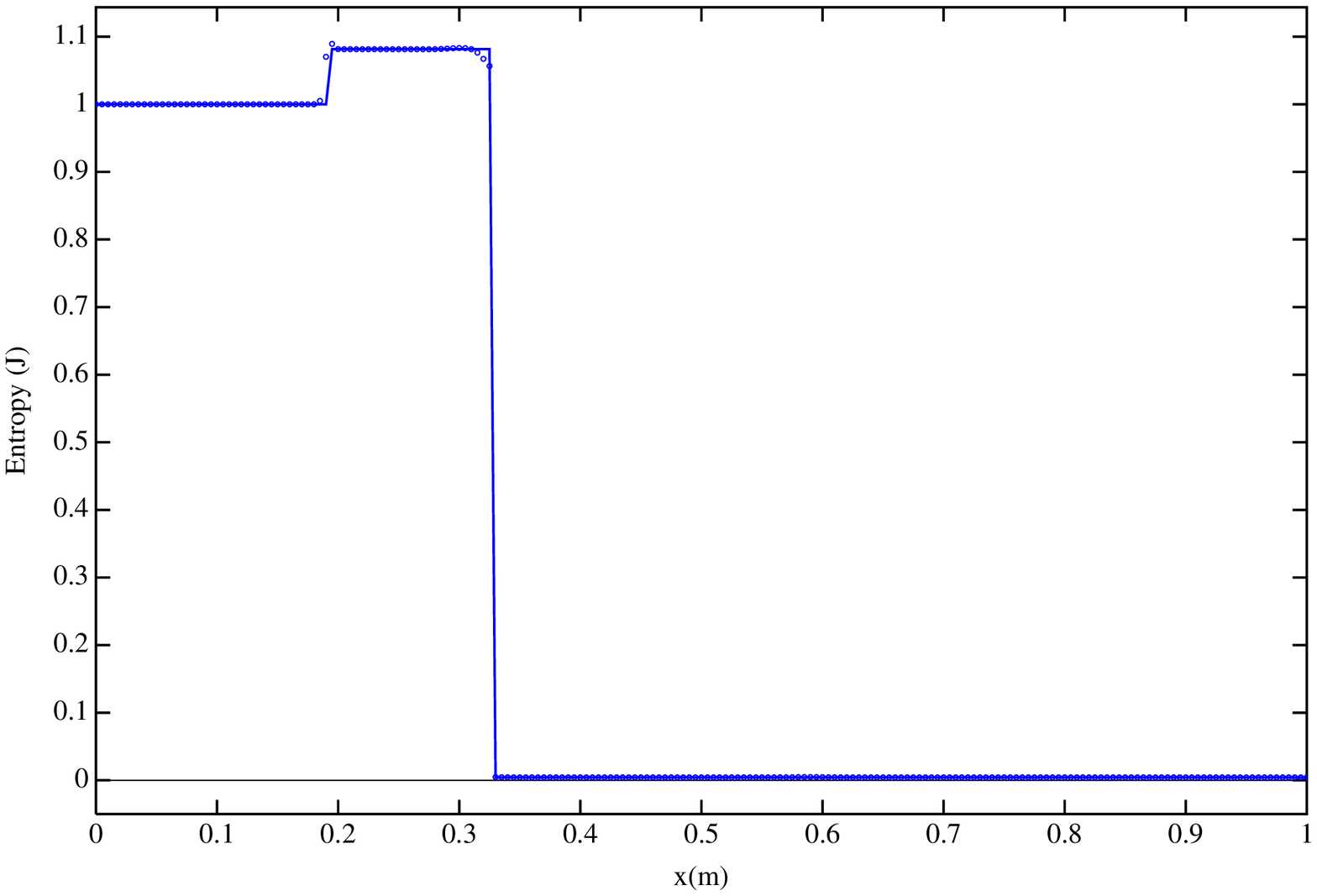}&  
\includegraphics[width=5.5cm,height=5.5cm]{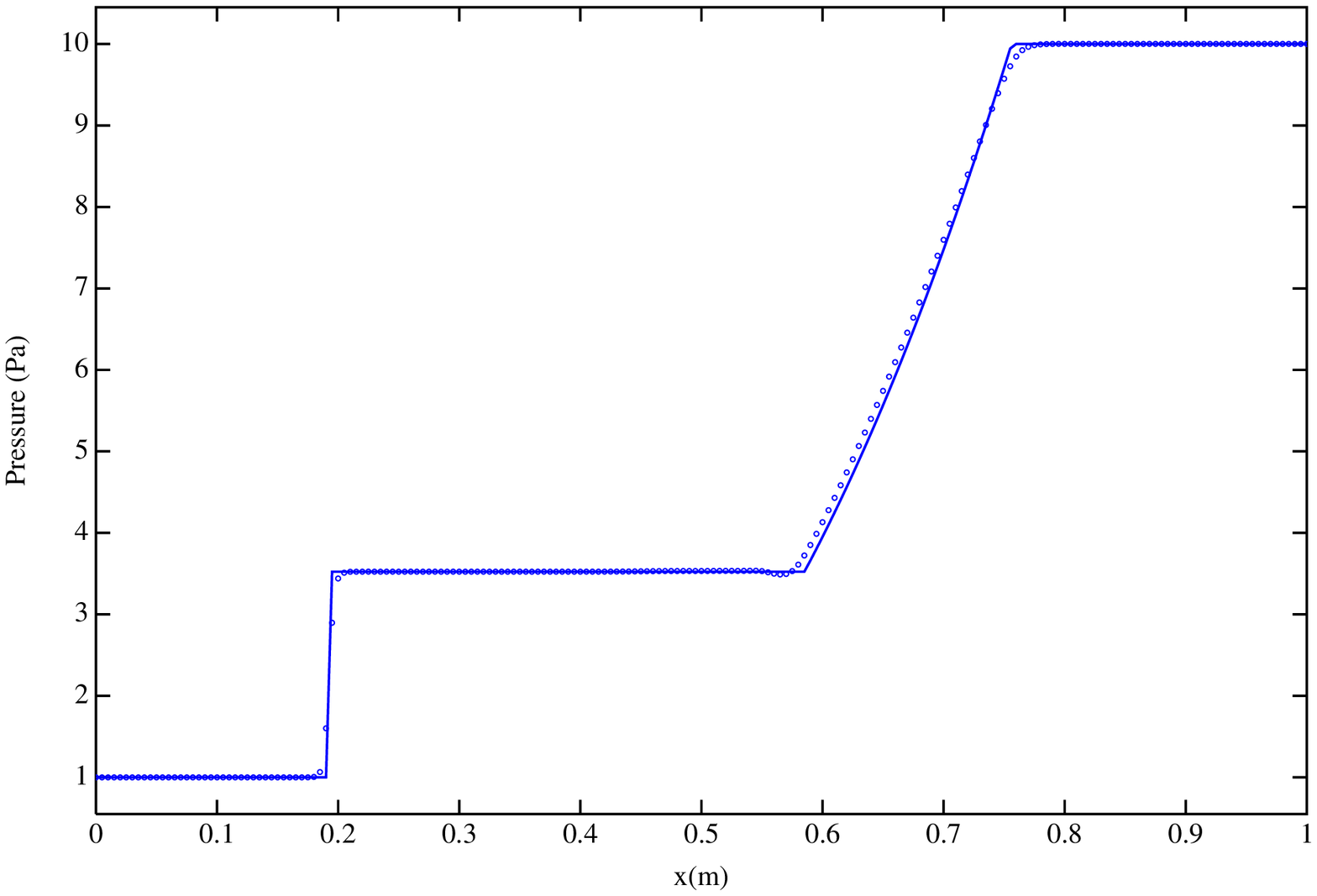} 
\end{tabular} 
\caption{Test 1-a: Mach 1.95 shock wave interacting with a material interface. WENO-5 coupled with the ESIM (exact values: solid line; numerical values: points).}                                                                                                                                                                                                                                                                                                                                                                                                                                                                                        \label{Test1a} 
\end{center} 
\end{figure} 
 
Firstly, we consider a shock-interface interaction problem, previously studied in \cite{SHYUE1}. On a 1 m long domain, the initial configuration consists of a stationary material interface at $\alpha_0=0.5$ m, separating two fluids with different EOS: a polytropic gas on the left, a stiffened gas on the right. A left-going Mach 1.95 shock wave is initially set at $\alpha_1=0.7$ m. Physical parameters are 
\begin{equation} 
\left\{ 
\begin{array}{l} 
\rho_0= 1.000\mbox{ kg/m}^3,\, p_0=1\mbox{ Pa },\, u_0=0\mbox{ m/s }, \gamma_0=1.4,\, p_{\infty\,0}=0\mbox{ Pa},\\ 
\\ 
\rho_1= 5.000\mbox{ kg/m}^3,\, p_1=1\mbox{ Pa },\, u_1=0\mbox{ m/s }, \gamma_1=4,\, p_{\infty\,1}=1\mbox{ Pa},\\ 
\\ 
\rho_2= 7.093\mbox{ kg/m}^3,\, p_2=10\mbox{ Pa }, u_2=-0.7288\mbox{ m/s }, \gamma_2=4.0,\, p_{\infty\,2}=1\mbox{ Pa}. 
\end{array} 
\right. 
\label{Test1_weak}
\end{equation} 
At $t=4.69\,10^{-2}$ s, the shock wave collides with the material interface, leading to a left-going shock, a left-going material interface, and a right-going rarefaction fan (see figure 4 in \cite{SHYUE1}). 
 
\begin{figure}[htbp] 
\begin{center} 
\begin{tabular}{cc} 
Density & Velocity\\ 
\includegraphics[width=5.5cm,height=5.5cm]{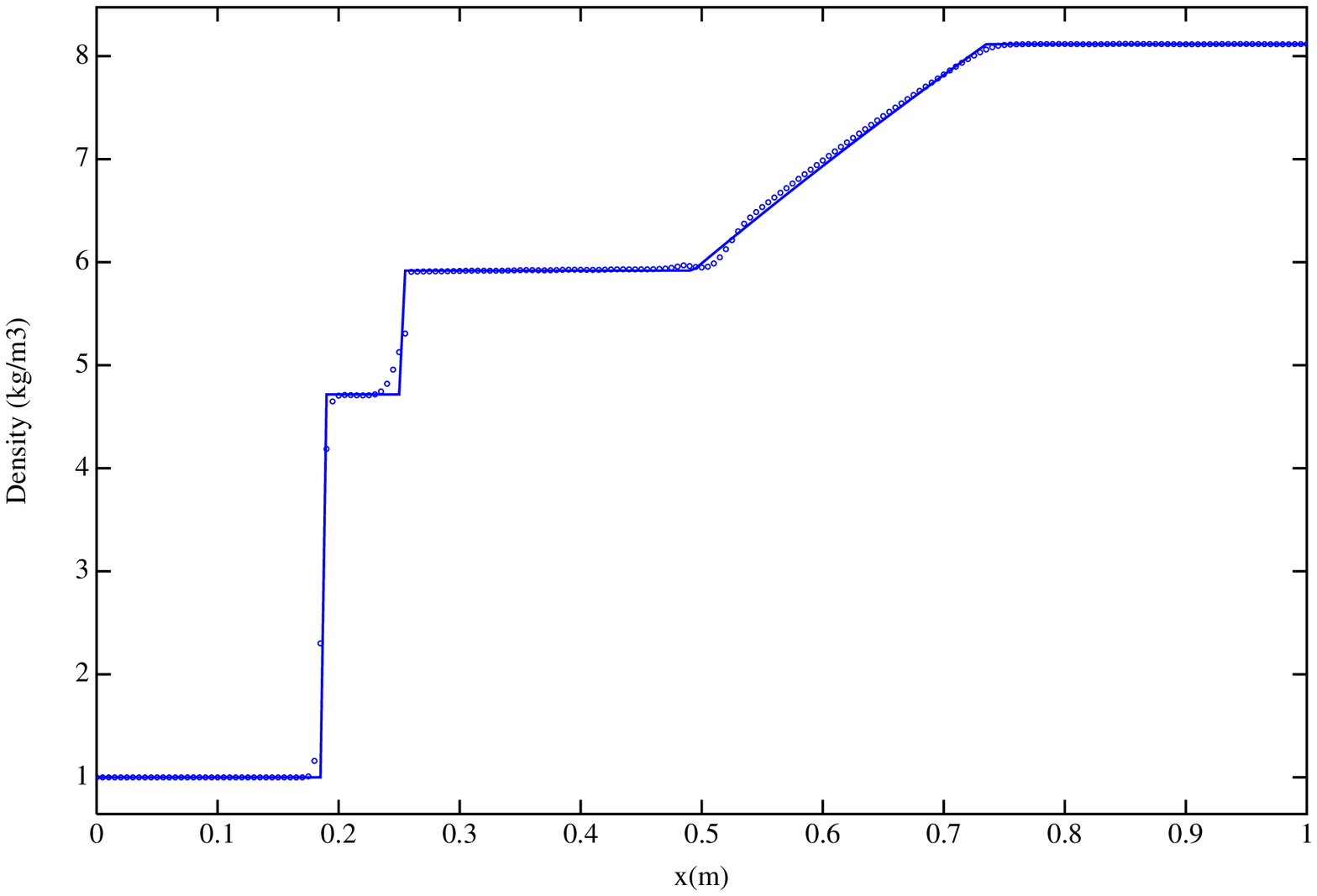}&  
\includegraphics[width=5.5cm,height=5.5cm]{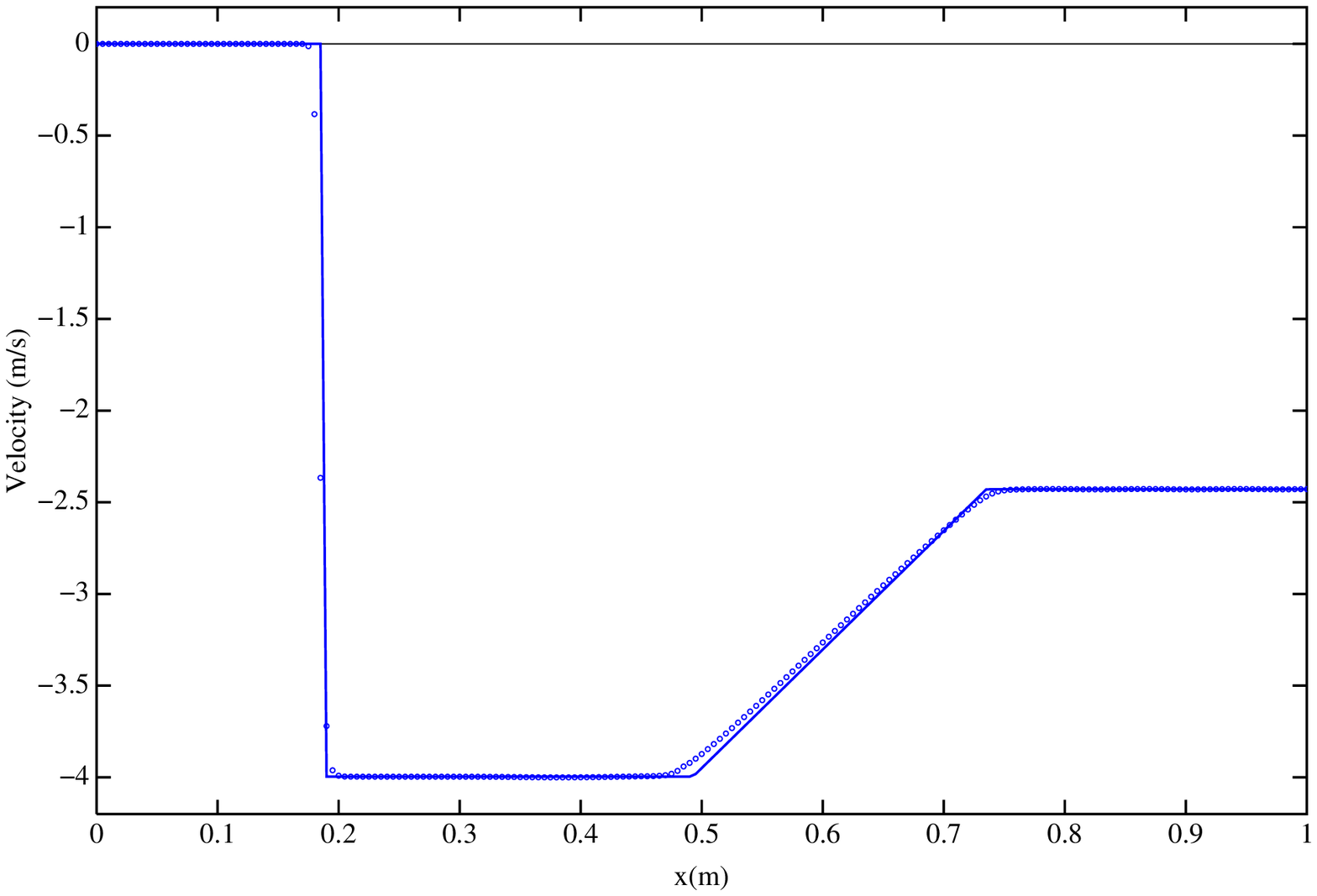}\\ 
Entropy & Pressure\\ 
\includegraphics[width=5.5cm,height=5.5cm]{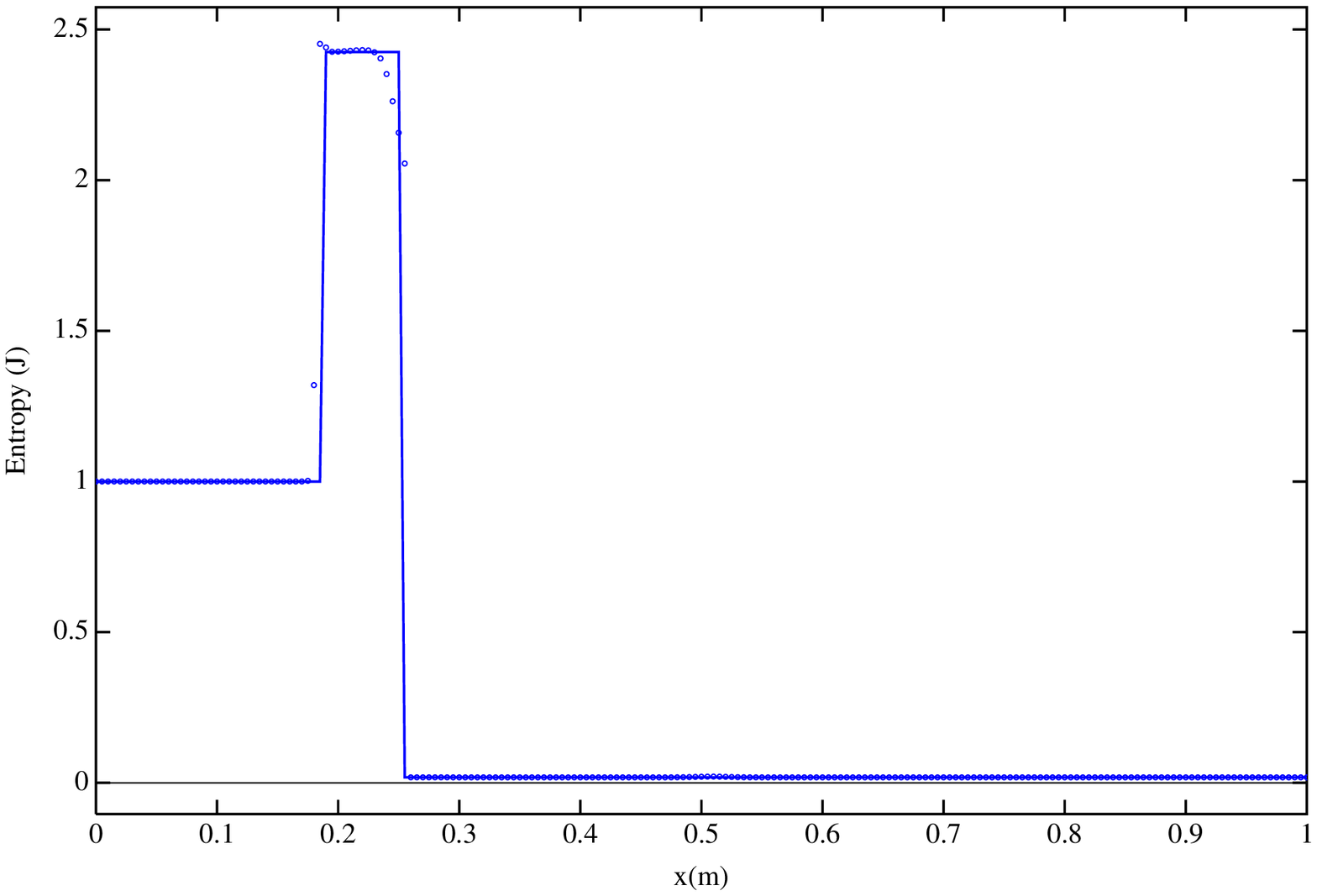}&  
\includegraphics[width=5.5cm,height=5.5cm]{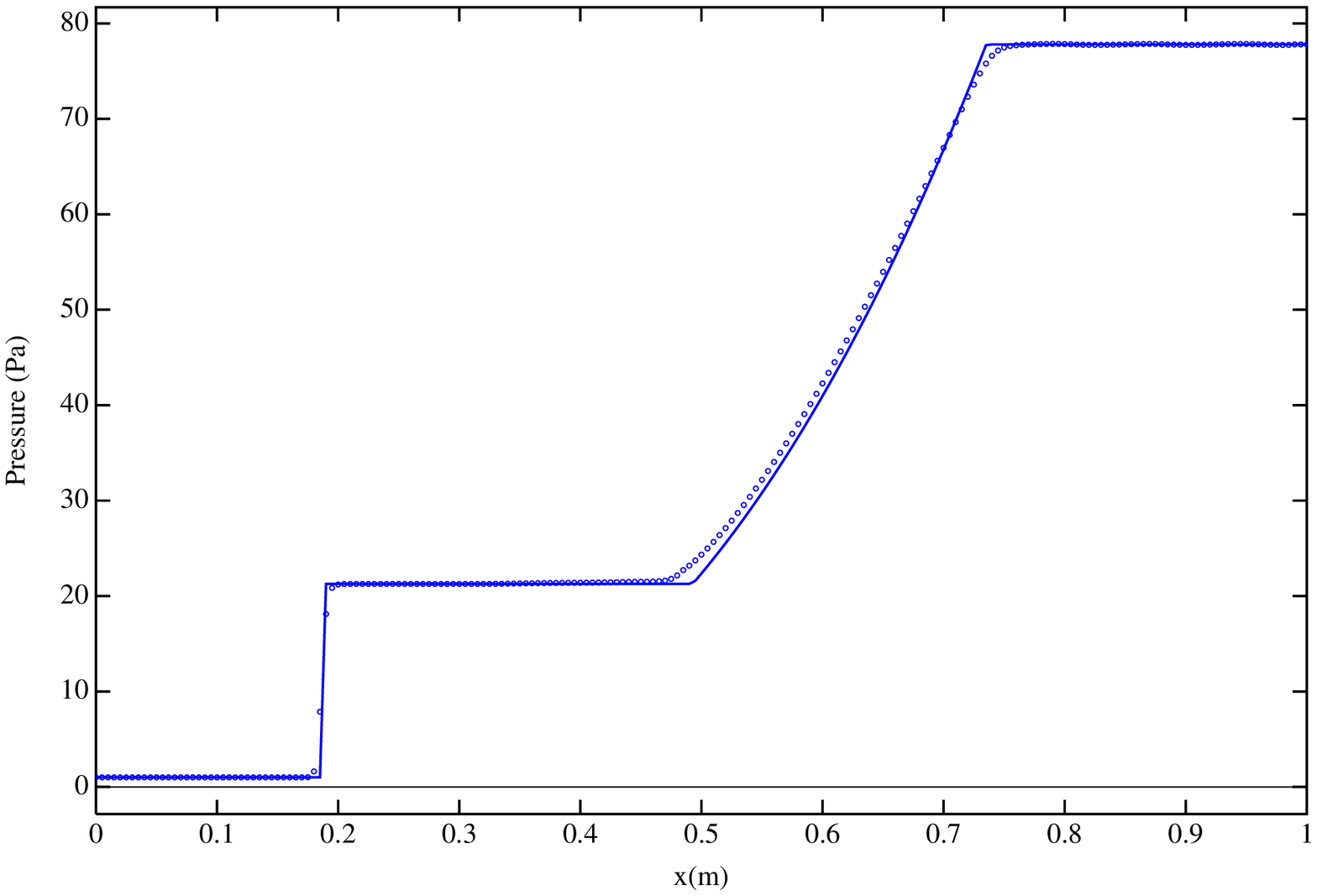} 
\end{tabular} 
\caption{Test 1-b: Mach 5 shock wave interacting with a material interface. WENO-5 coupled with the ESIM (exact values: solid line; numerical values: points).}                                                                                                                                                                                                                                                                                                                                                                                                                                                                                        \label{Test1b} 
\end{center} 
\end{figure} 

Figure \ref{Test1a} shows exact values and numerical values of $\rho$, $u$, $p$, and entropy $S$ at $t=0.202$ s (after 200 time steps). The computations are performed on $N_x$ = 200 grid points with WENO-5 coupled with the ESIM. The agreement between analytical and numerical values is good. The main interest of this example is to show that the ESIM is robust and behaves well, even when a shock wave is in the vicinity of the material interface. No spurious oscillations induced by Taylor expansions are seen in $u$ and $p$ around the material interface (near $x=0.32$ m).

Secondly, we consider the case of a stronger shock wave. The physical parameters around the material interface are the same than in (\ref{Test1_weak}); only the parameters in the post-shock fluid are modified
$$
\rho_2= 8.116\mbox{ kg/m}^3,\quad p_2=77.80\mbox{ Pa },\quad u_2=-2.428\mbox{ m/s },\quad \gamma_2=1.4,\quad p_{\infty\,2}=1\mbox{ Pa}, 
$$
what amounts to a Mach 5 left-going shock wave. The wave phenomena (after the collision between the shock wave and the material interface) are the same than in the previous example. The computations are performed with WENO-5 coupled with the ESIM. 

Figure \ref{Test1b} shows exact values and numerical values of $\rho$, $u$, $p$, and entropy $S$ at $t=0.112$ s (after 300 time steps). The agreement between numerical and analytical values is good, even for this strong shock test. No oscillations are visible in $u$ and $p$ around the material interface (near $x=0.26$ m). 

\subsection{Test 2: shock-bubble interaction} 
 
\begin{figure}[htbp] 
\begin{center} 
\begin{tabular}{cc} 
Density & Velocity\\ 
\includegraphics[width=5.5cm,height=5.5cm]{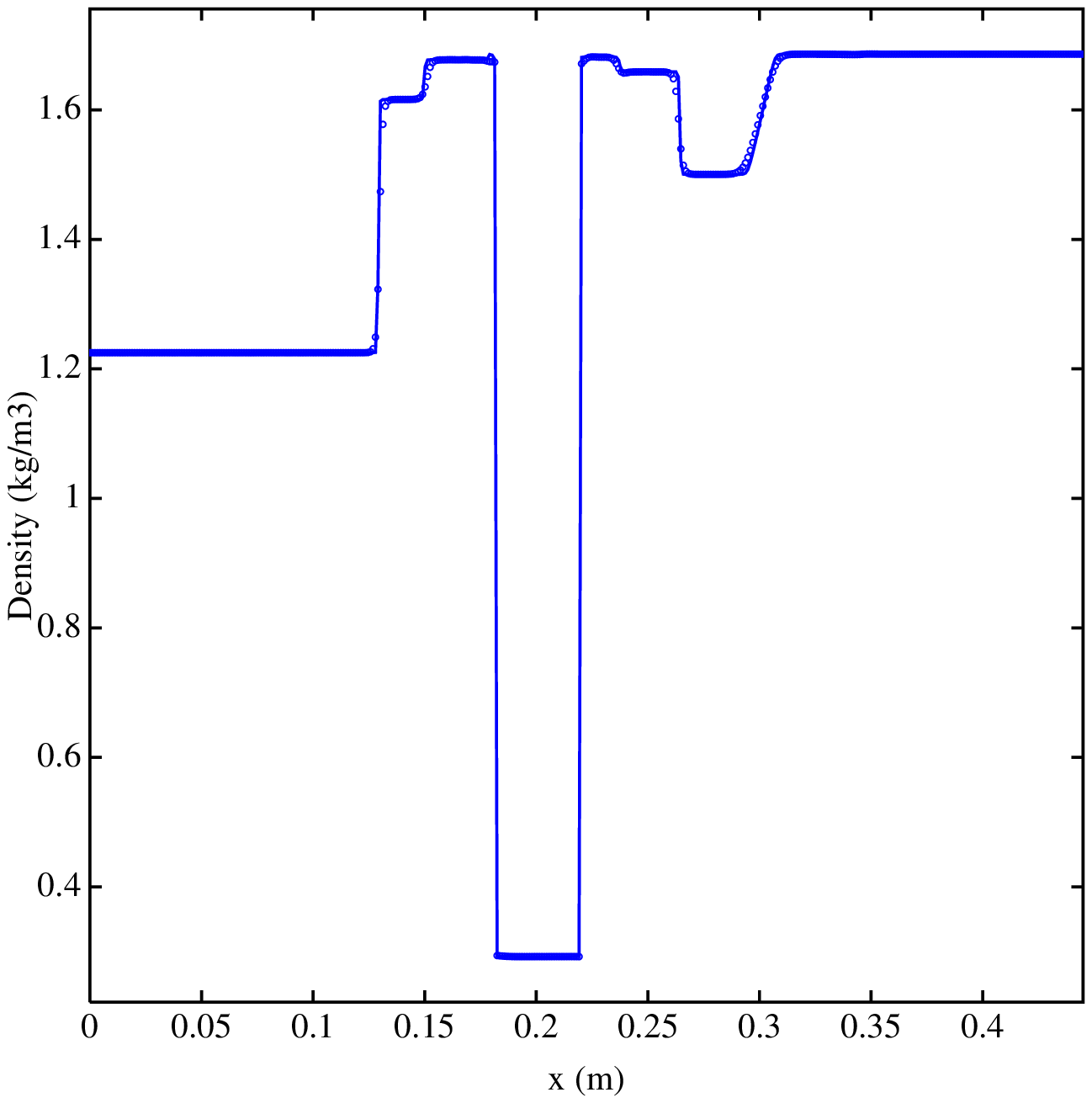}&  
\includegraphics[width=5.5cm,height=5.5cm]{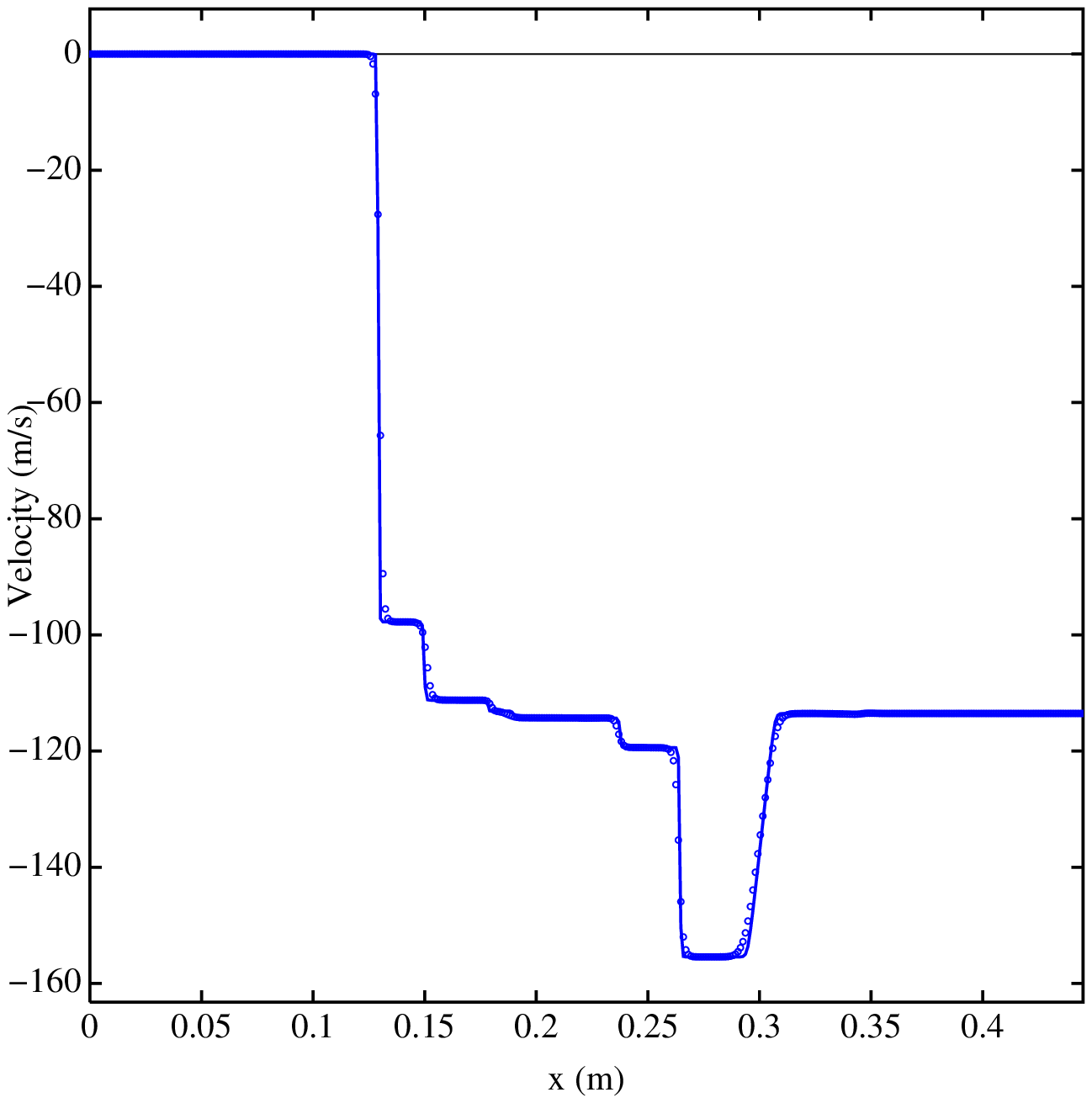}\\ 
Entropy & Pressure\\ 
\includegraphics[width=5.5cm,height=5.5cm]{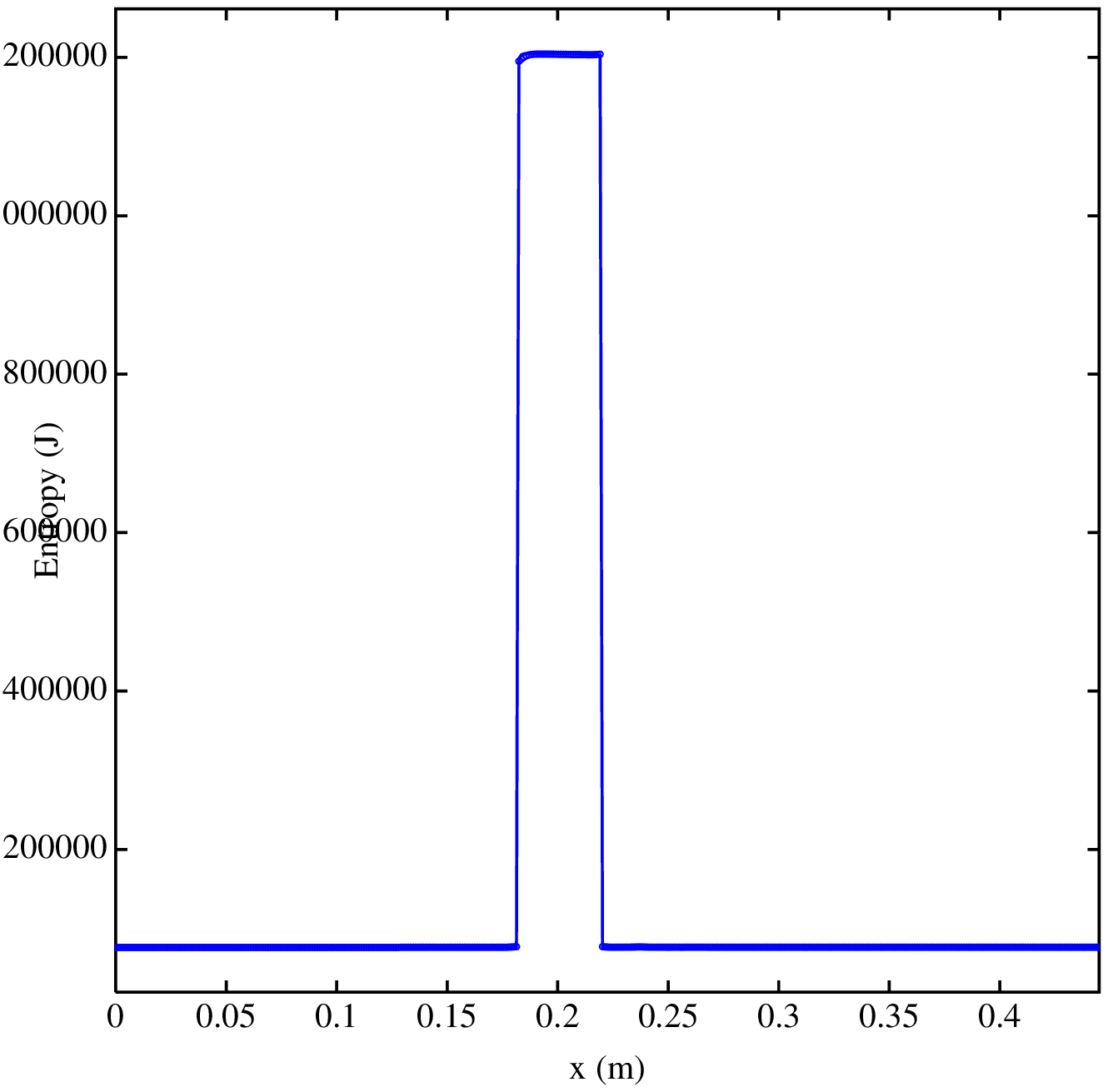}&  
\includegraphics[width=5.5cm,height=5.5cm]{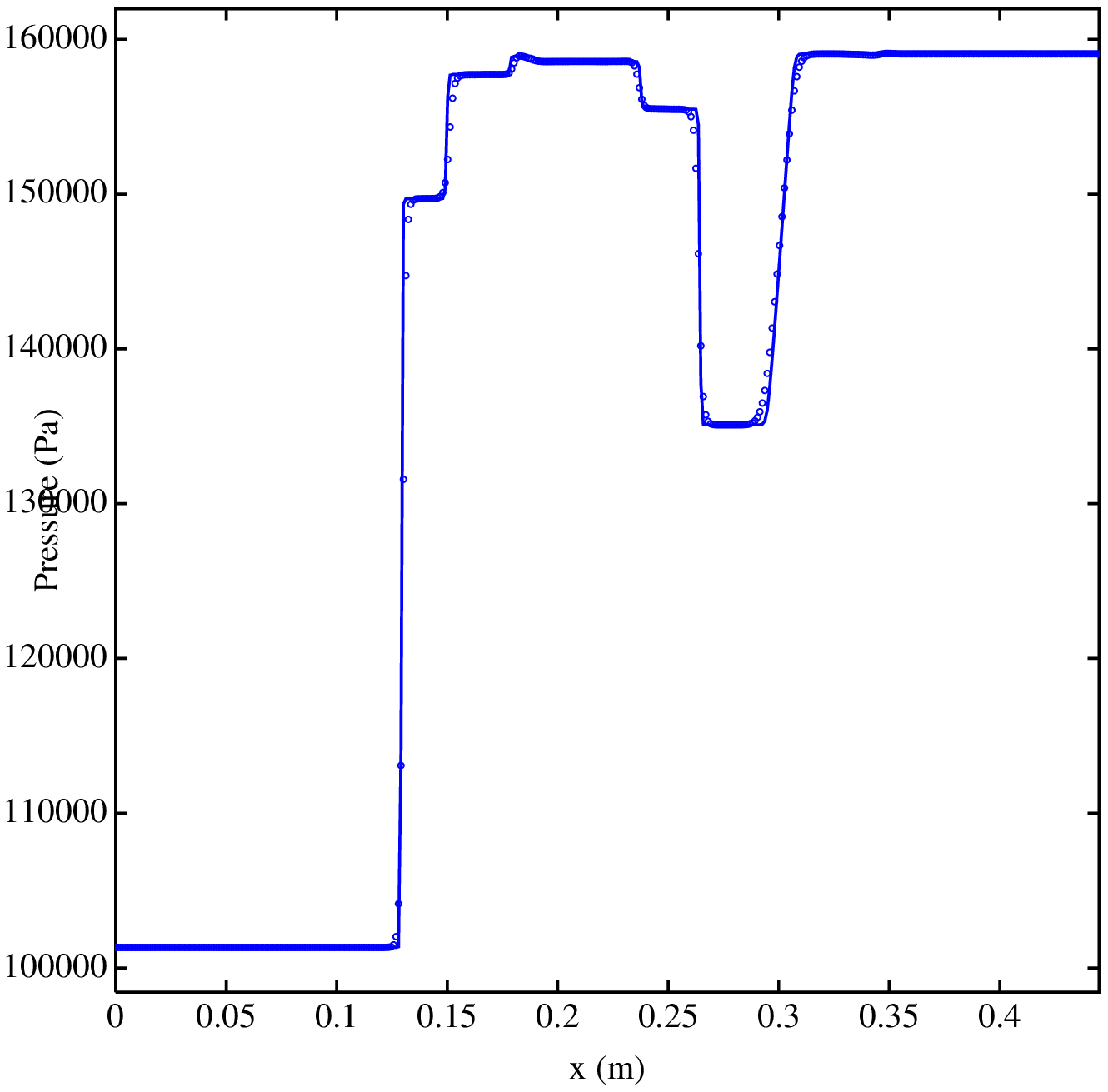} 
\end{tabular} 
\caption{Test 2: shock-bubble interaction. WENO-5 coupled with the ESIM (fine grid solution: solid line; numerical values: points).}                                                                                                                                                                                                                                                                                                                                                                                                                                                                                        
\label{Test2} 
\end{center} 
\end{figure} 
We consider the 1D version of the classical shock-bubble interaction problem, numerically addressed e.g. in \cite{MULET} with a conservative algorithm. On a 0.445 m long domain, a helium domain, initially delimitated by $\alpha_0=0.2$ m and $\alpha_1=0.25$ m, is at rest in air. A left-going Mach 1.22 shock wave is initially set at $\alpha_2=0.275$ m. Physical parameters are 
$$ 
\left\{ 
\begin{array}{l} 
\rho_0= 1225\mbox{ kg/m}^3,\, p_0=1.01325\,10^5\mbox{ Pa },\, u_0=0.0\mbox{ m/s }, \gamma_0=1.4,\, p_{\infty\,0}=0\mbox{ Pa},\\ 
\\ 
\rho_1= 0.2228\mbox{ kg/m}^3,\, p_1=1.01325\,10^5\mbox{ Pa },\, u_1=0.0\mbox{ m/s }, \gamma_1=1.648,\, p_{\infty\,1}=0\mbox{ Pa},\\ 
\\ 
\rho_2= 1225\mbox{ kg/m}^3,\, p_2=1.01325\,10^5\mbox{ Pa },\, u_2=0.0\mbox{ m/s }, \gamma_2=1.4,\, p_{\infty\,2}=0\mbox{ Pa},\\ 
\\ 
\rho_3= 1686\mbox{ kg/m}^3,\, p_3=1.59059\,10^5\mbox{ Pa },\, u_3=-3.59\mbox{ m/s }, \gamma_3=1.4,\, p_{\infty\,3}=0\mbox{ Pa.} 
\end{array} 
\right. 
$$ 
See \cite{MULET} for a description of wave phenomena. Since no analytic solution is available, we compute the solution with WENO-5 coupled with ESIM on a fine grid with 3200 mesh points; we refer to this approximation as the "exact solution" for comparison purposes. 
 
Figure \ref{Test2} shows "exact values" and numerical values of $\rho$, $u$, $p$, and $S$ at $t=2.86\,10^{-3}$ s (after 400 time steps). The computations are performed on $N_x$ = 400 grid points with WENO-5 coupled to the ESIM. The agreement between "exact values" and numerical values is very good. The resolution is better than in \cite{MULET}: compare the density inside the helium in figure \ref{Test2} with the density in figure 2 of \cite{MULET}. 

To conclude the tests 1 and 2, let us notice that similar results could be obtained with the GFM treatment (not shown here). Indeed, when only flat profiles are involved, the GFM and the ESIM have a similar behavior. When a shock wave collides with a material interface, both treatments produce a "sloping" behavior in density and entropy very close to the "overheating" phenomenon found in shock reflection problems \cite{DONAT}. Increasing the Mach number seems to accentuate this "overheating", producing in addition slightly perturbed post-shock values: small acoustic perturbations can be observed downstream in figure \ref{Test1b}. More testing (not shown) has been performed with Mach numbers up to 9, which confirm these observations. No crash due to negative pressure values has been observed on these tests.

The goal of the next two tests is to show the advantage of using the ESIM (compared to the GFM) for rich-structured flows.
 
\subsection{Test 3: pure advection} 
 
\begin{figure}[htbp] 
\begin{center} 
\begin{tabular}{cc} 
Density (GFM) & Density (ESIM)\\ 
\includegraphics[width=5.0cm,height=5.0cm]{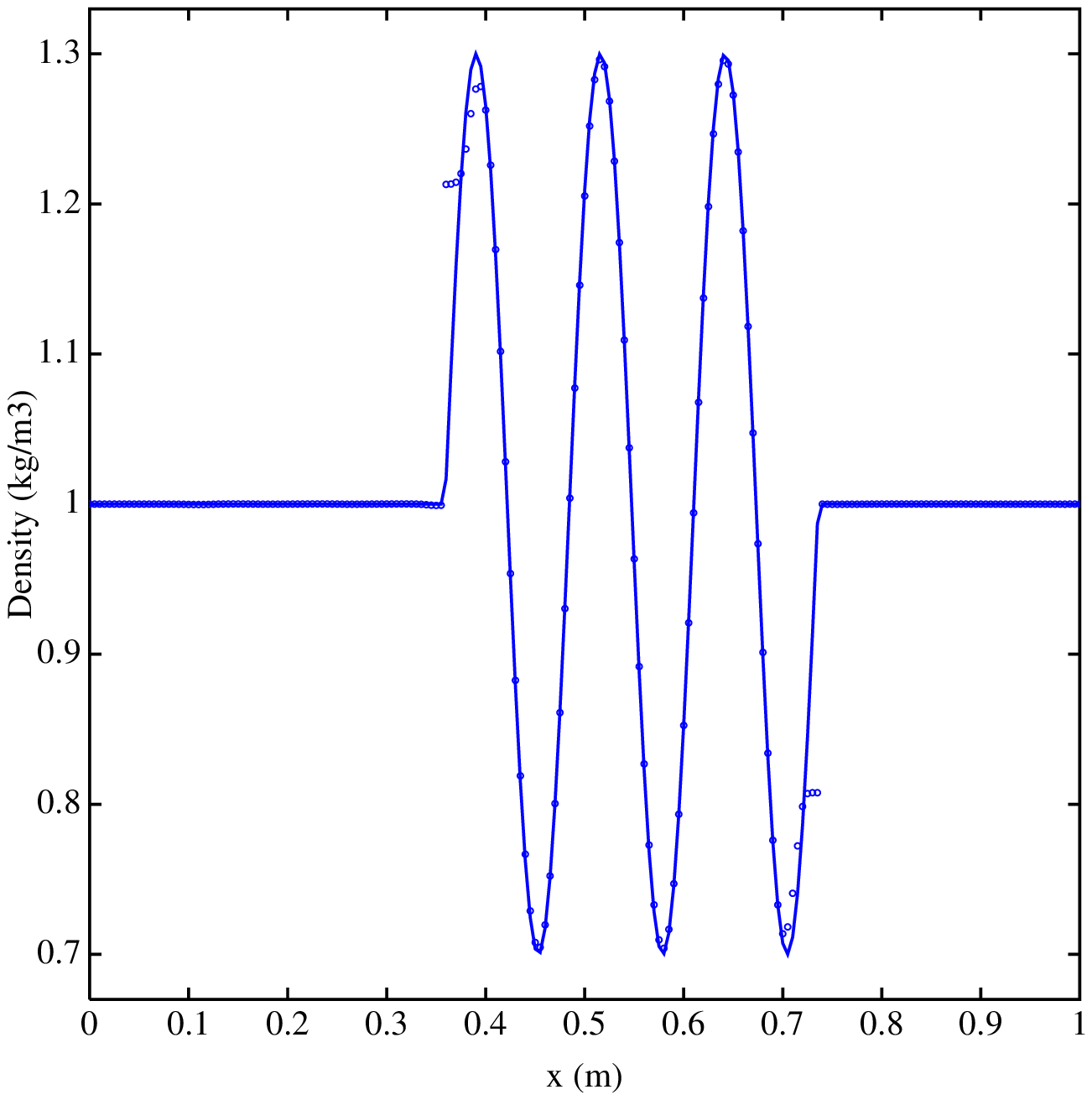}&  
\includegraphics[width=5.0cm,height=5.0cm]{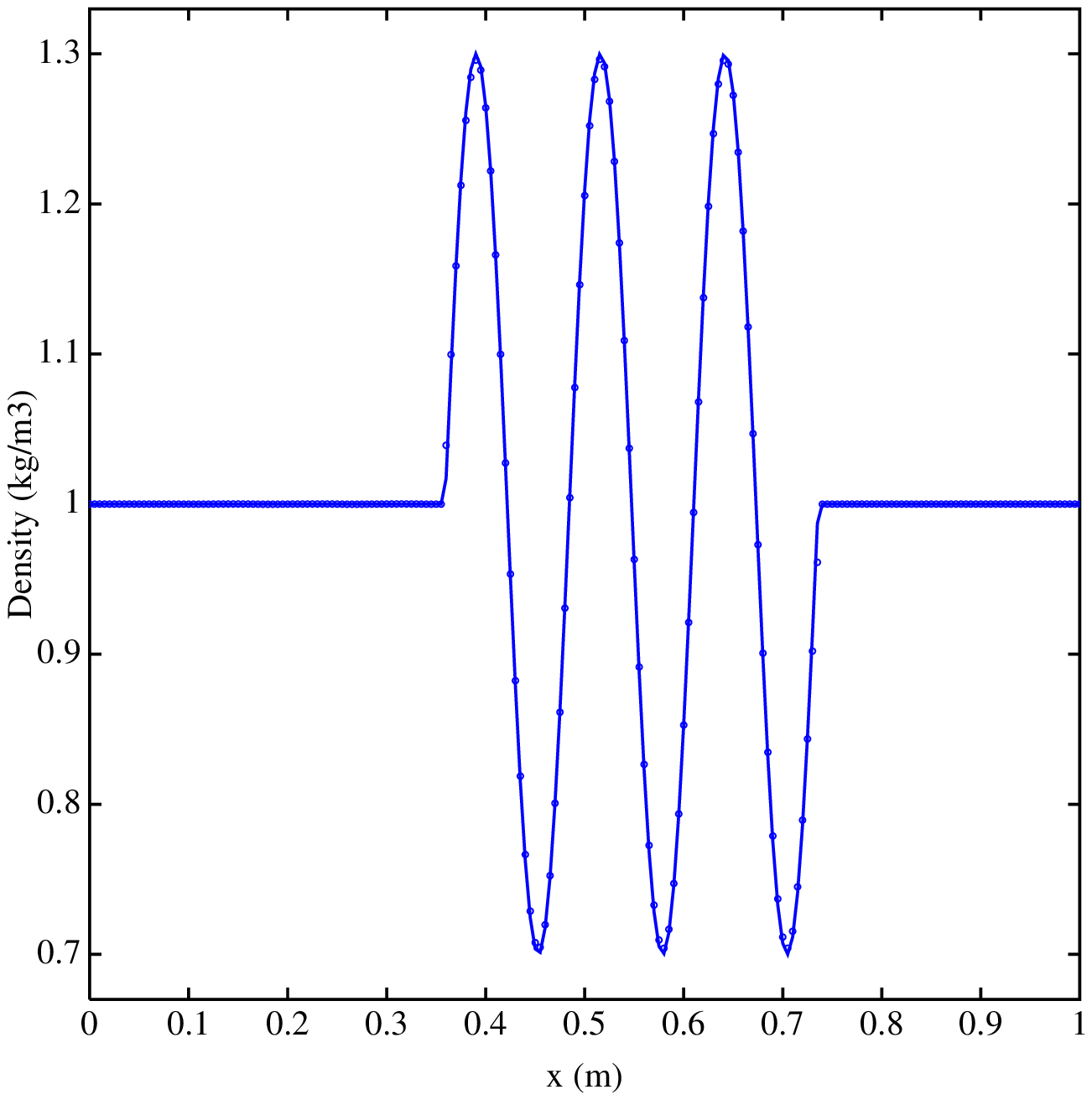}\\ 
Entropy (GFM) & Entropy (ESIM)\\ 
\includegraphics[width=5.0cm,height=5.0cm]{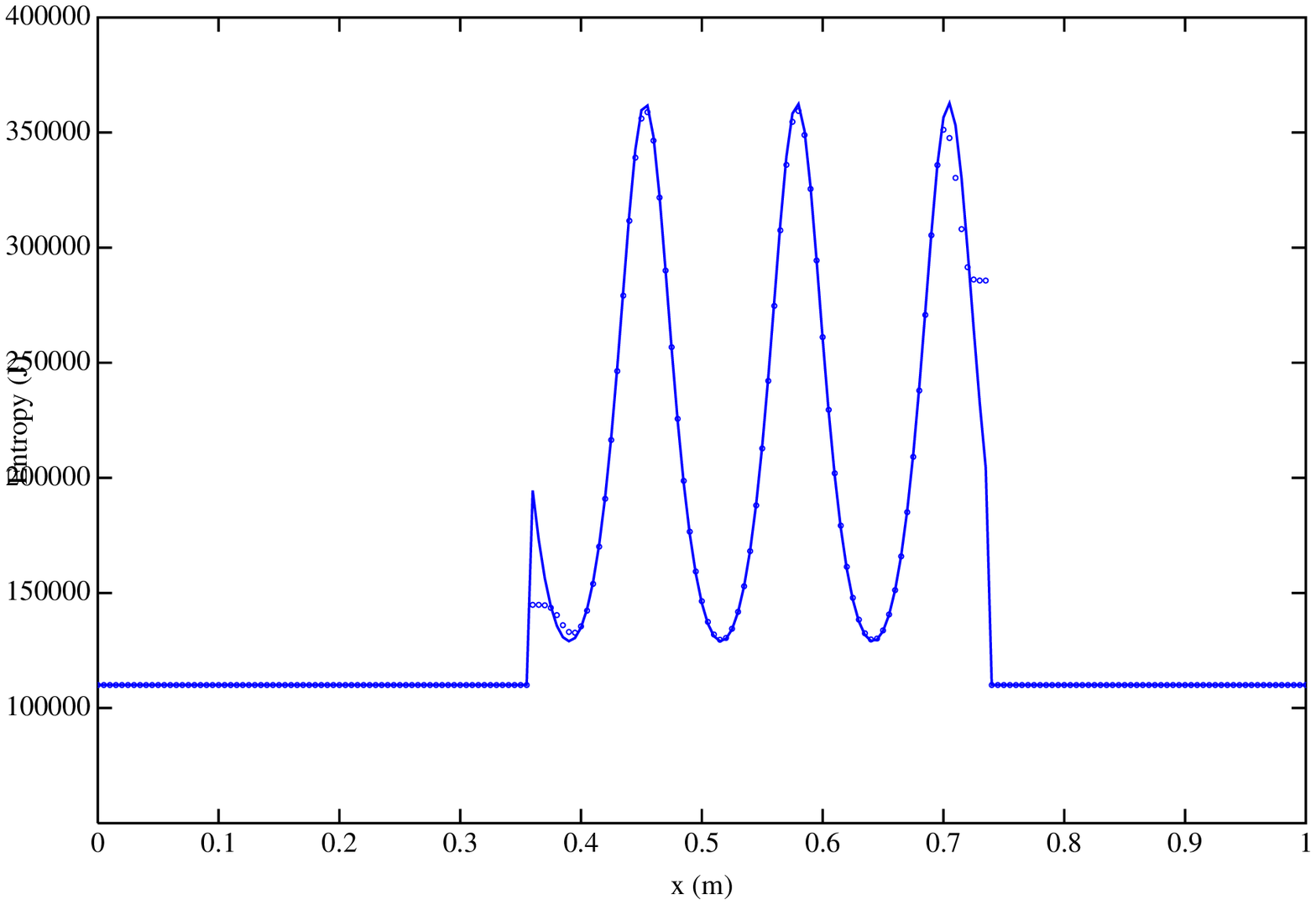}&  
\includegraphics[width=5.0cm,height=5.0cm]{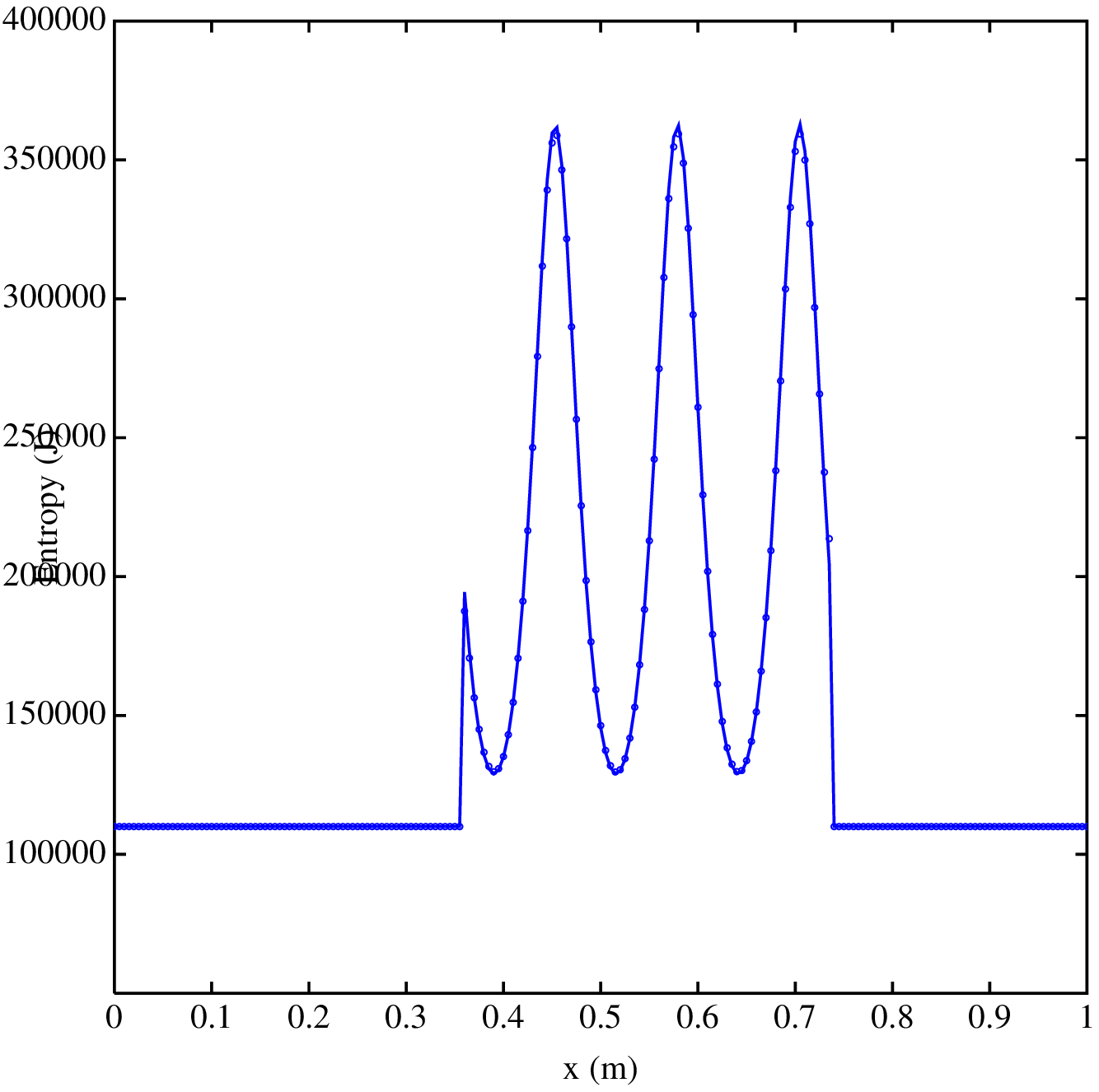}\\ 
Pressure (GFM) & Pressure (ESIM)\\ 
\includegraphics[width=5.0cm,height=5.0cm]{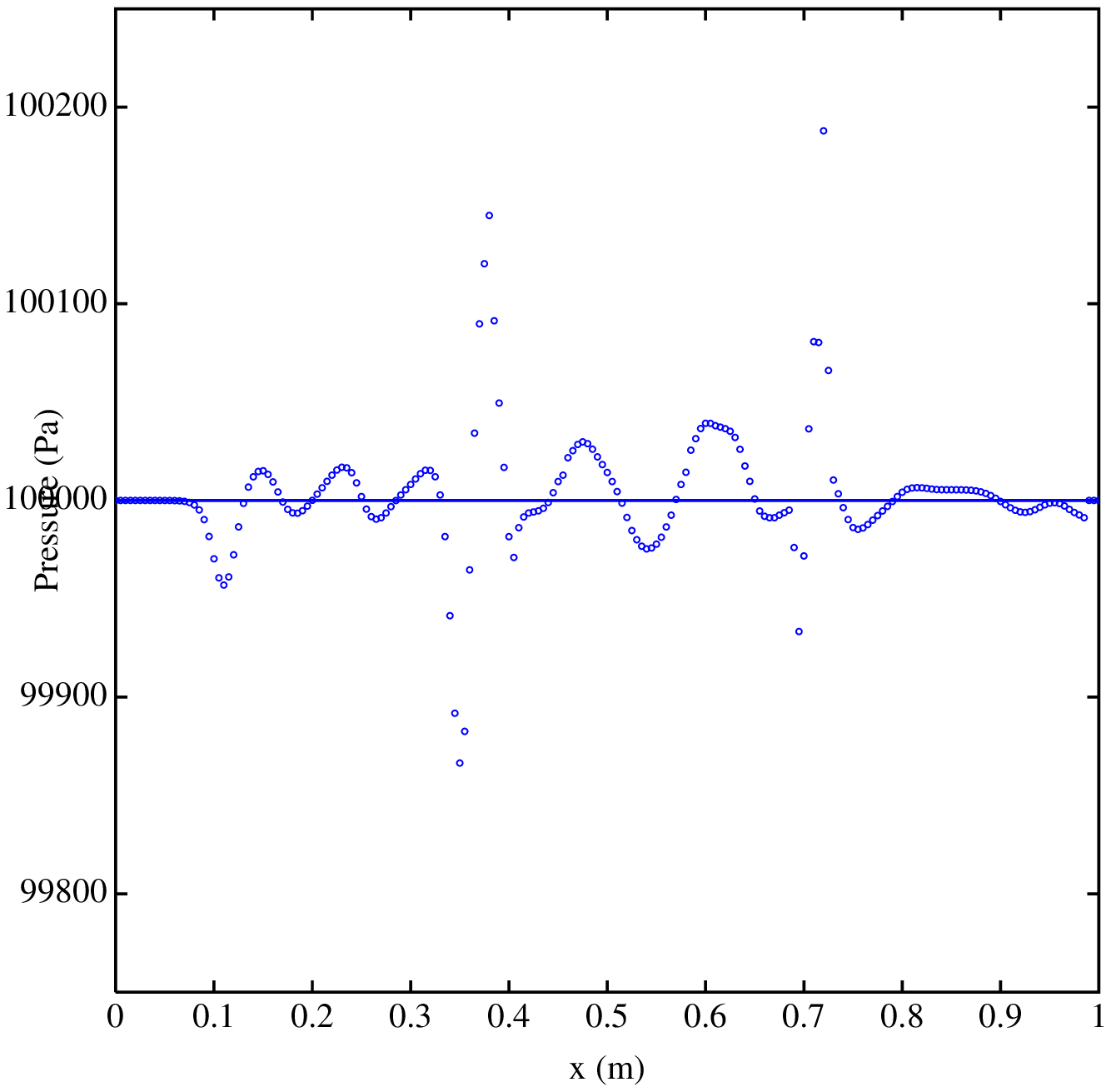}&  
\includegraphics[width=5.0cm,height=5.0cm]{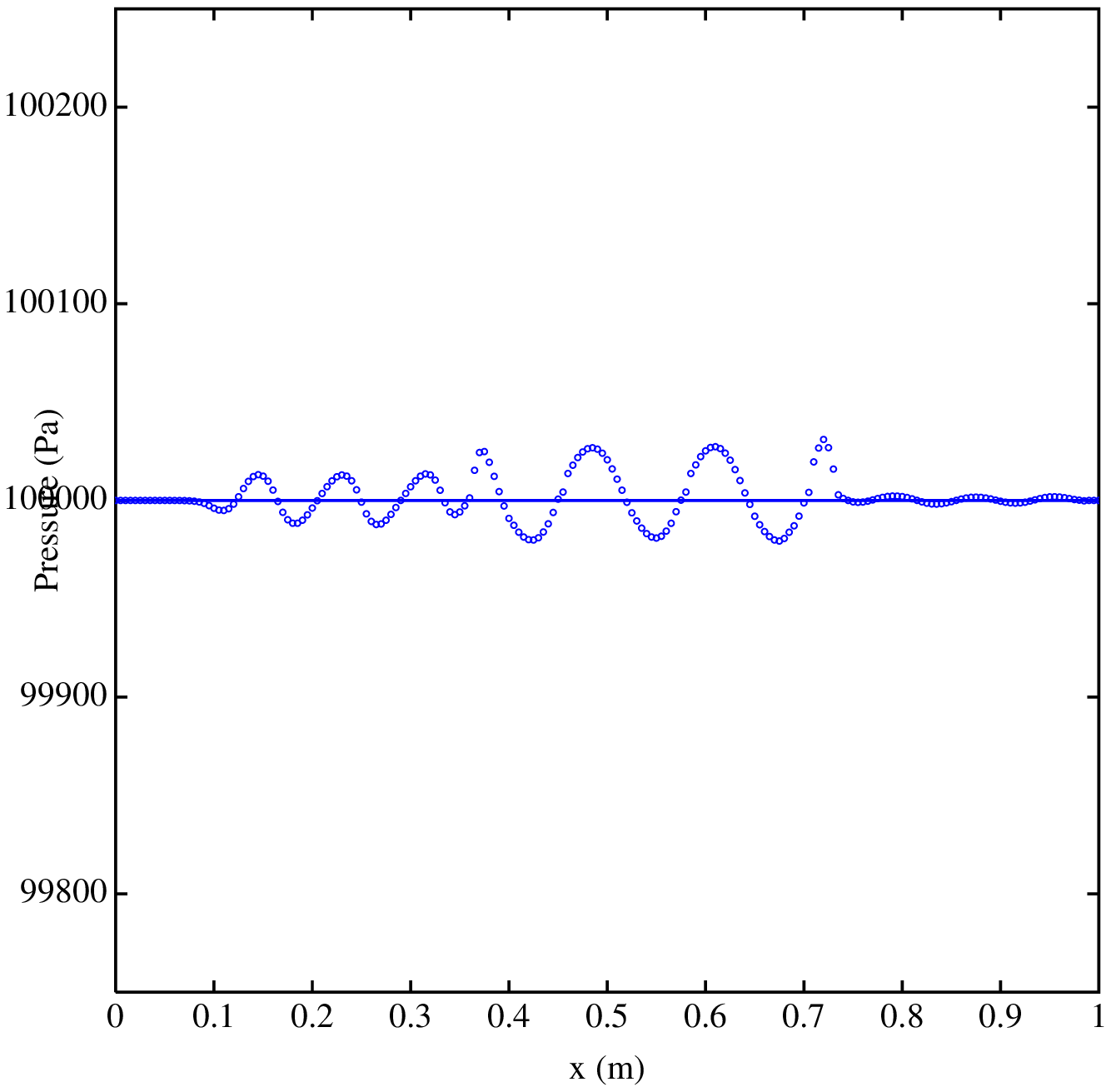}\\ 
Velocity (GFM) & Velocity (ESIM)\\ 
\includegraphics[width=5.0cm,height=5.0cm]{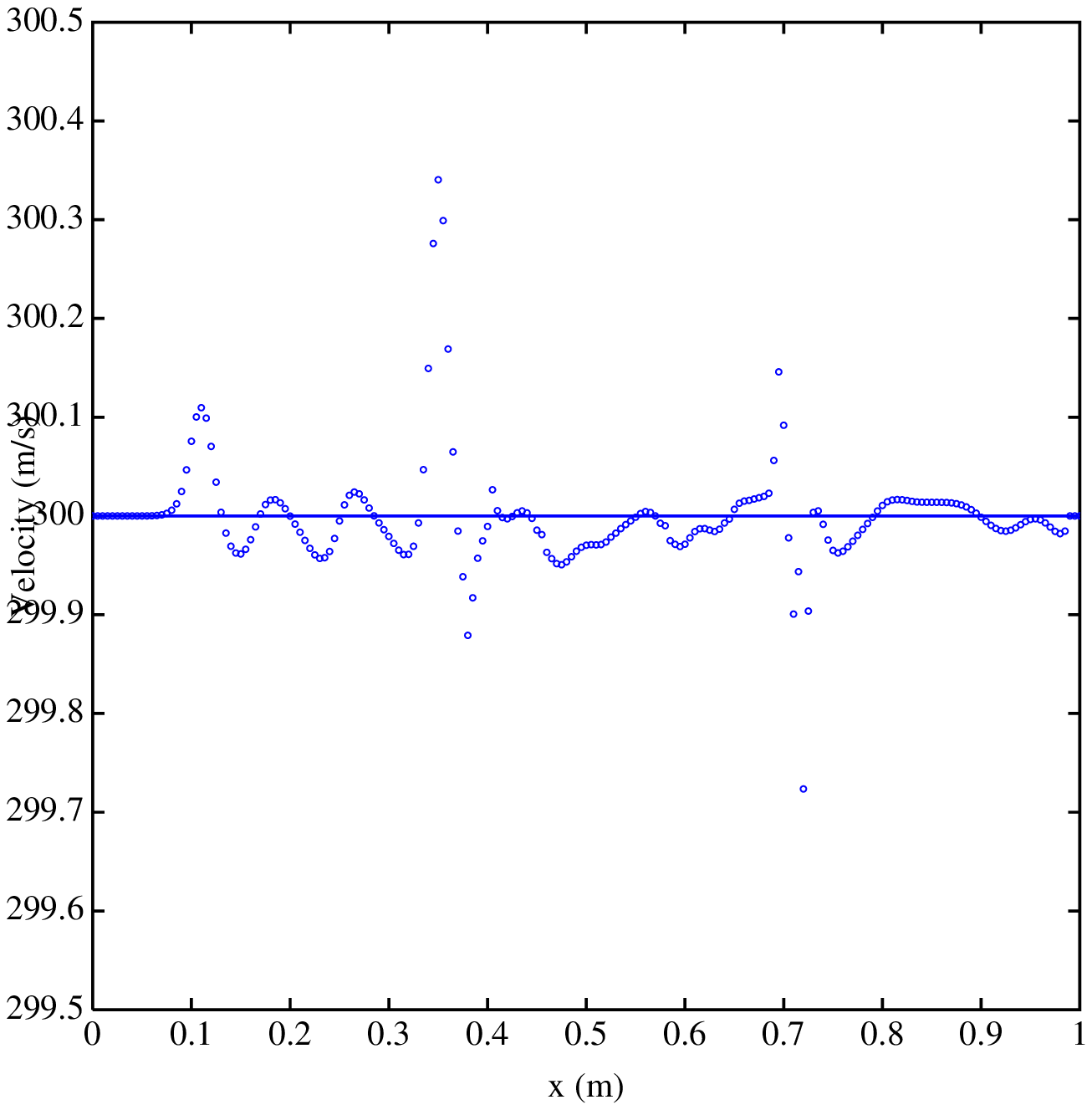}&  
\includegraphics[width=5.0cm,height=5.0cm]{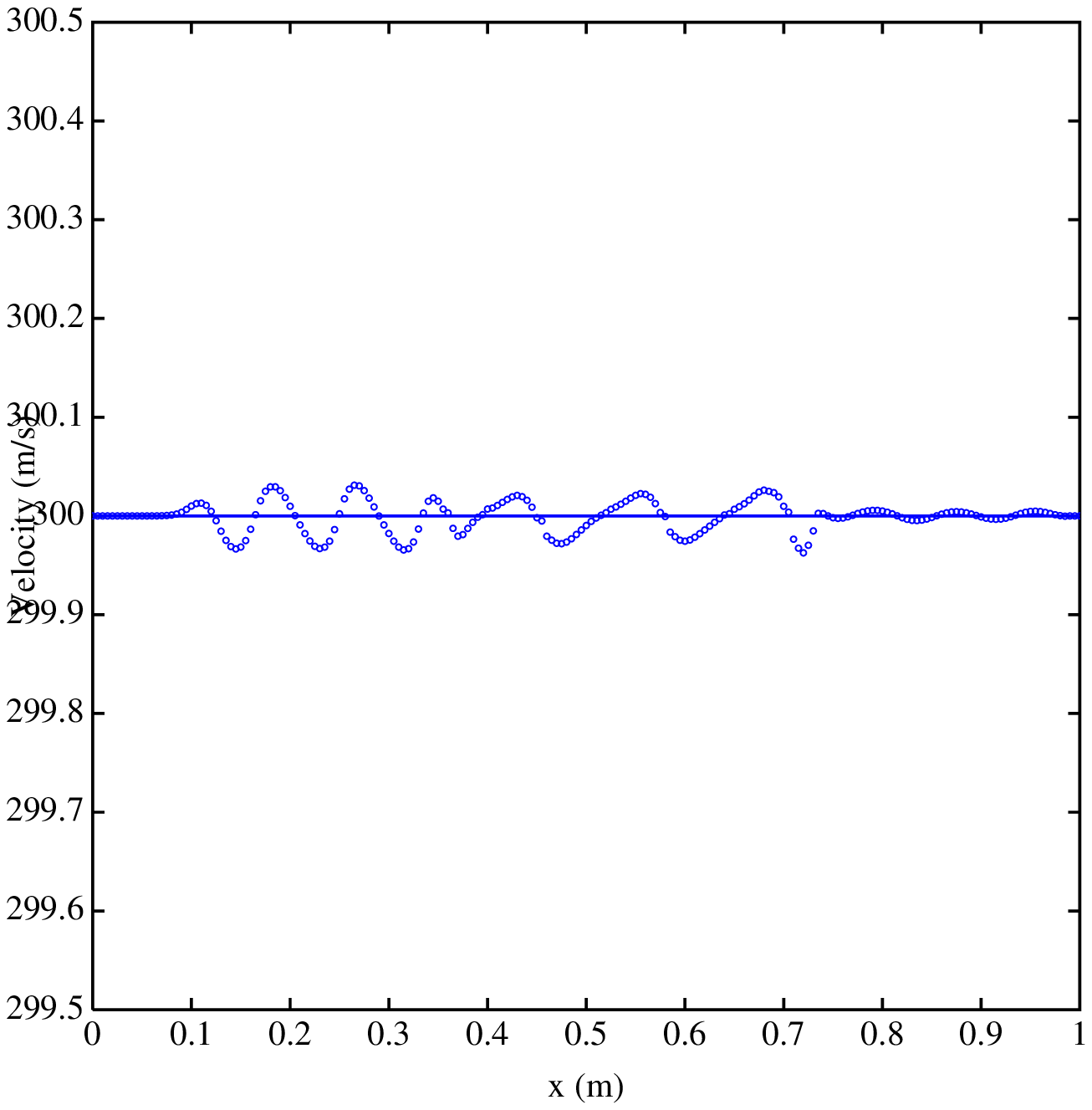} 
\end{tabular} 
\caption{Test 3-a: 200 grid points. ENO-3 coupled with the GFM (left column) and with the ESIM (right column). Exact values: solid line; numerical values: points (note that the scales for $p$ and $u$ are magnified).}                                                                                                                                                                                                                                                                                                                                                                                                                                                                                         
\label{Test3_200} 
\end{center} 
\end{figure} 
 
\begin{figure}[htbp] 
\begin{center} 
\begin{tabular}{cc} 
Density (GFM) & Density (ESIM)\\ 
\includegraphics[width=5.0cm,height=5.0cm]{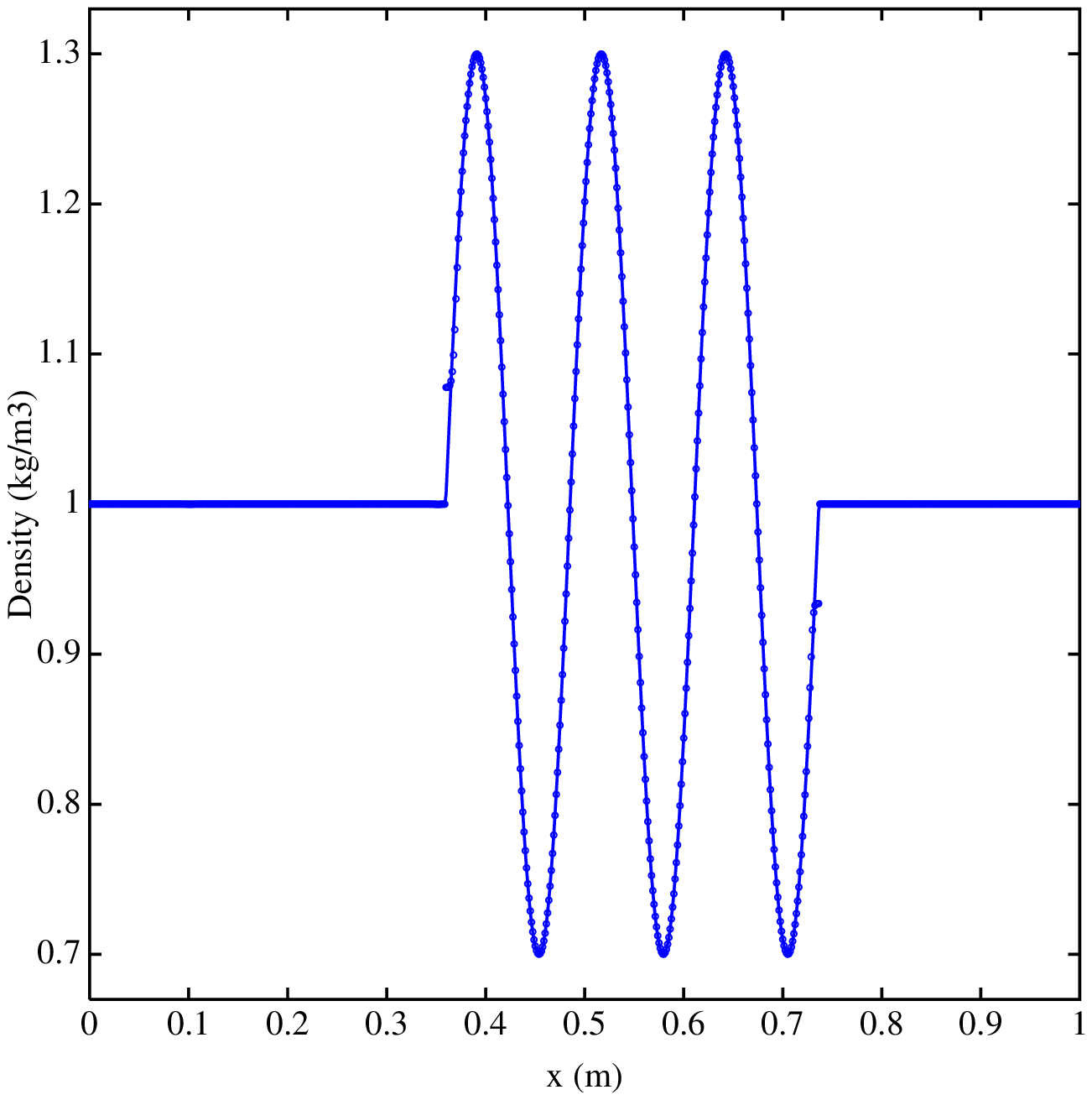}&  
\includegraphics[width=5.0cm,height=5.0cm]{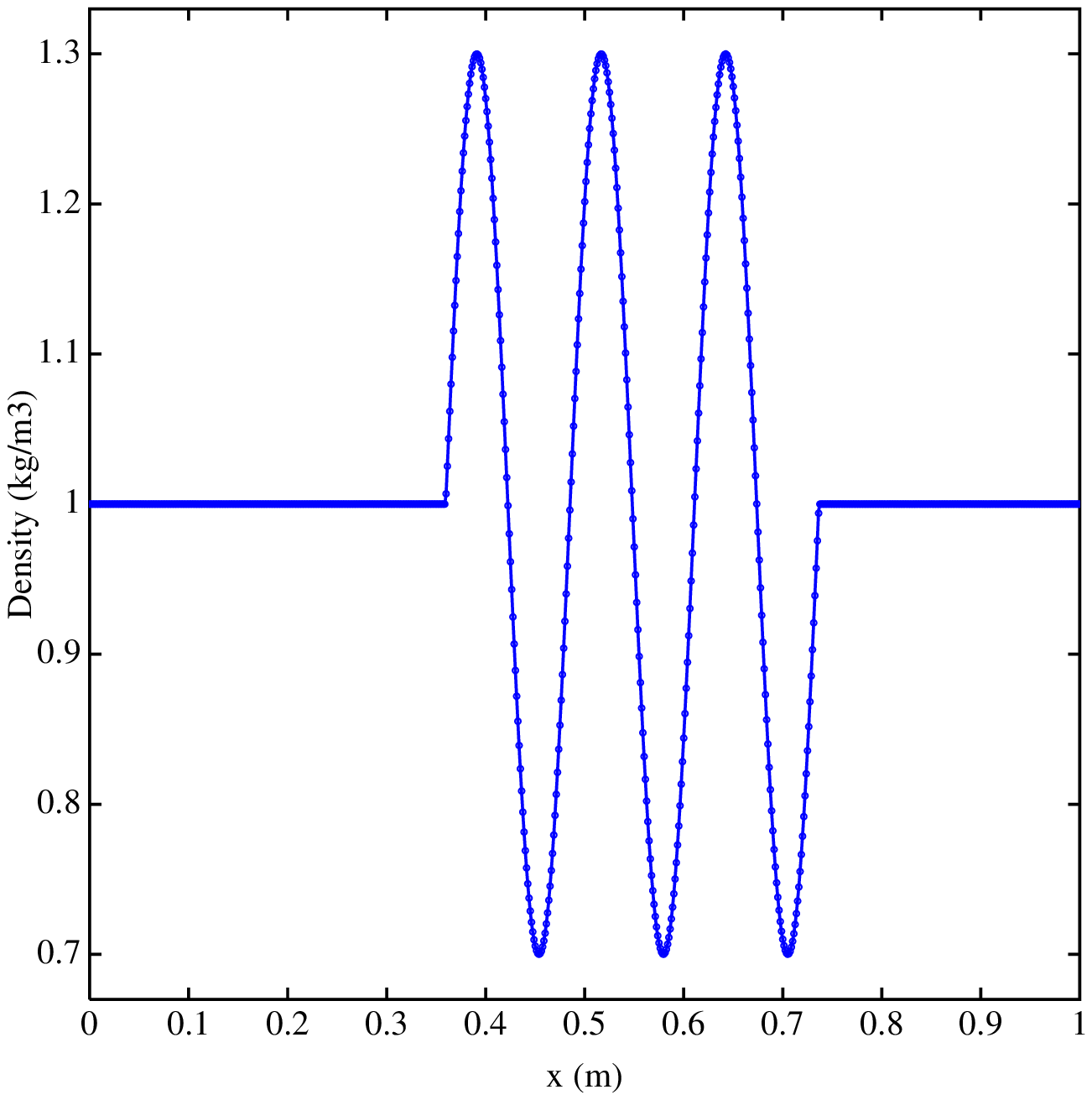}\\ 
Entropy (GFM) & Entropy (ESIM)\\ 
\includegraphics[width=5.0cm,height=5.0cm]{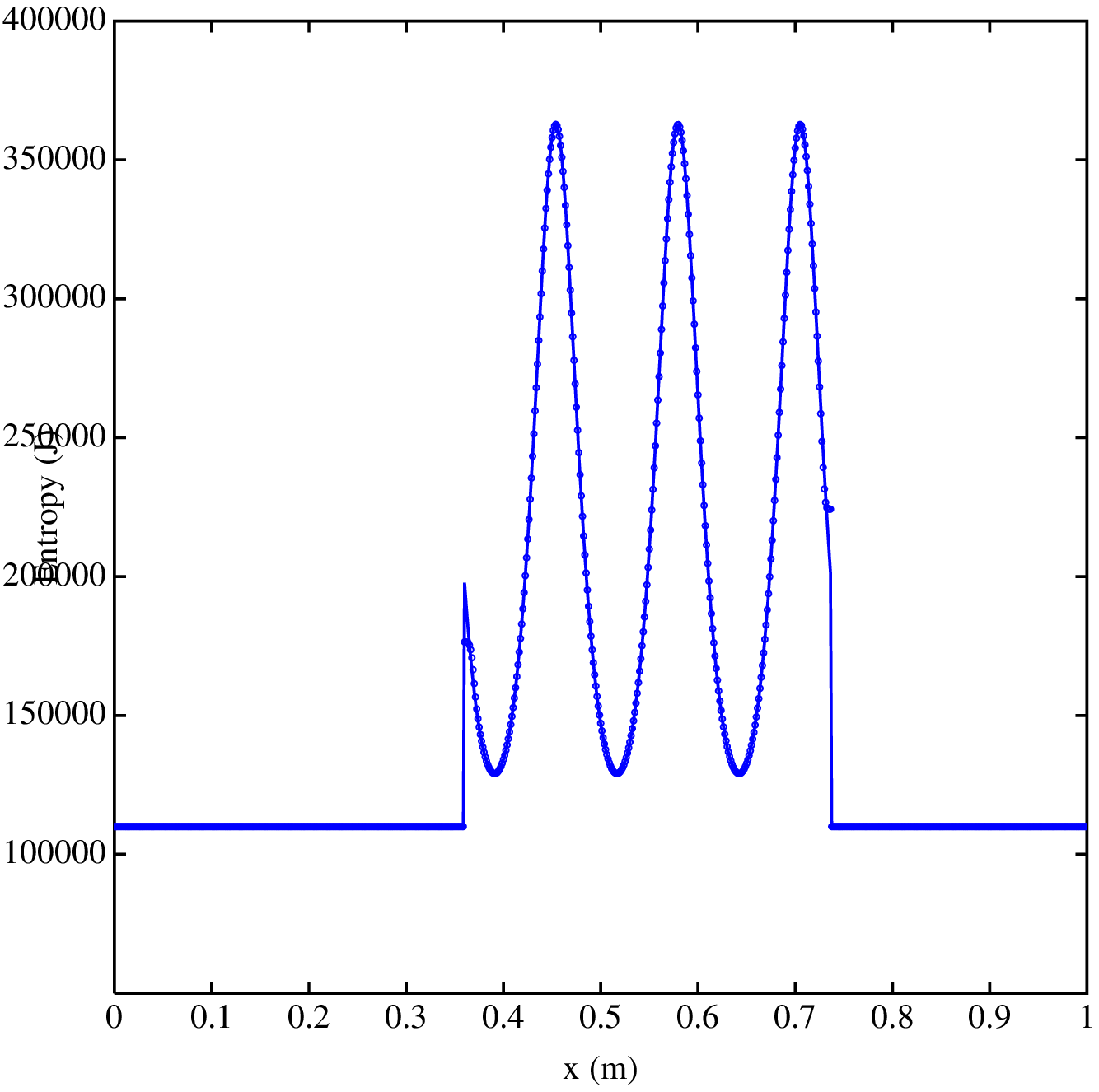}&  
\includegraphics[width=5.0cm,height=5.0cm]{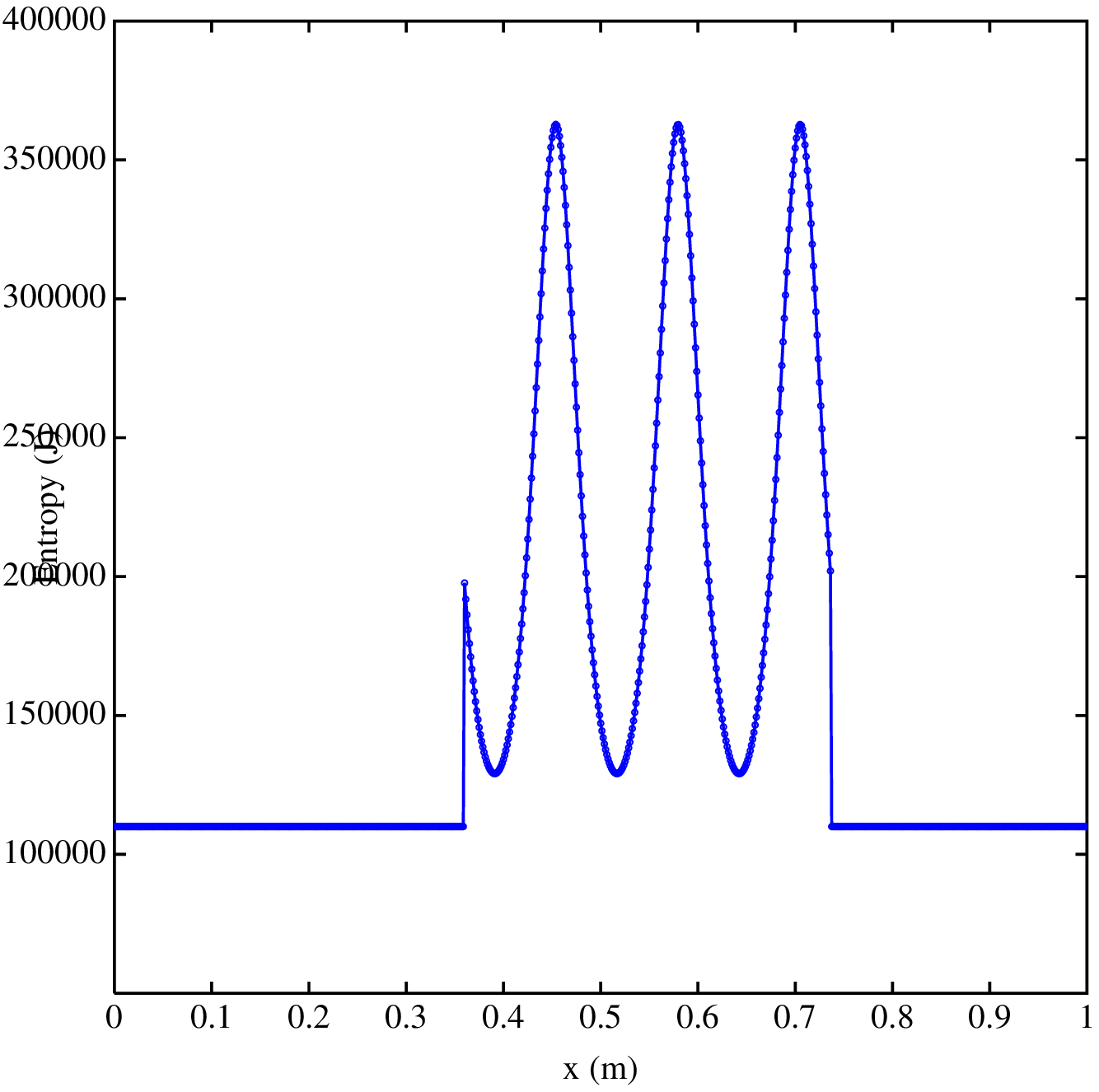}\\ 
Pressure (GFM) & Pressure (ESIM)\\ 
\includegraphics[width=5.0cm,height=5.0cm]{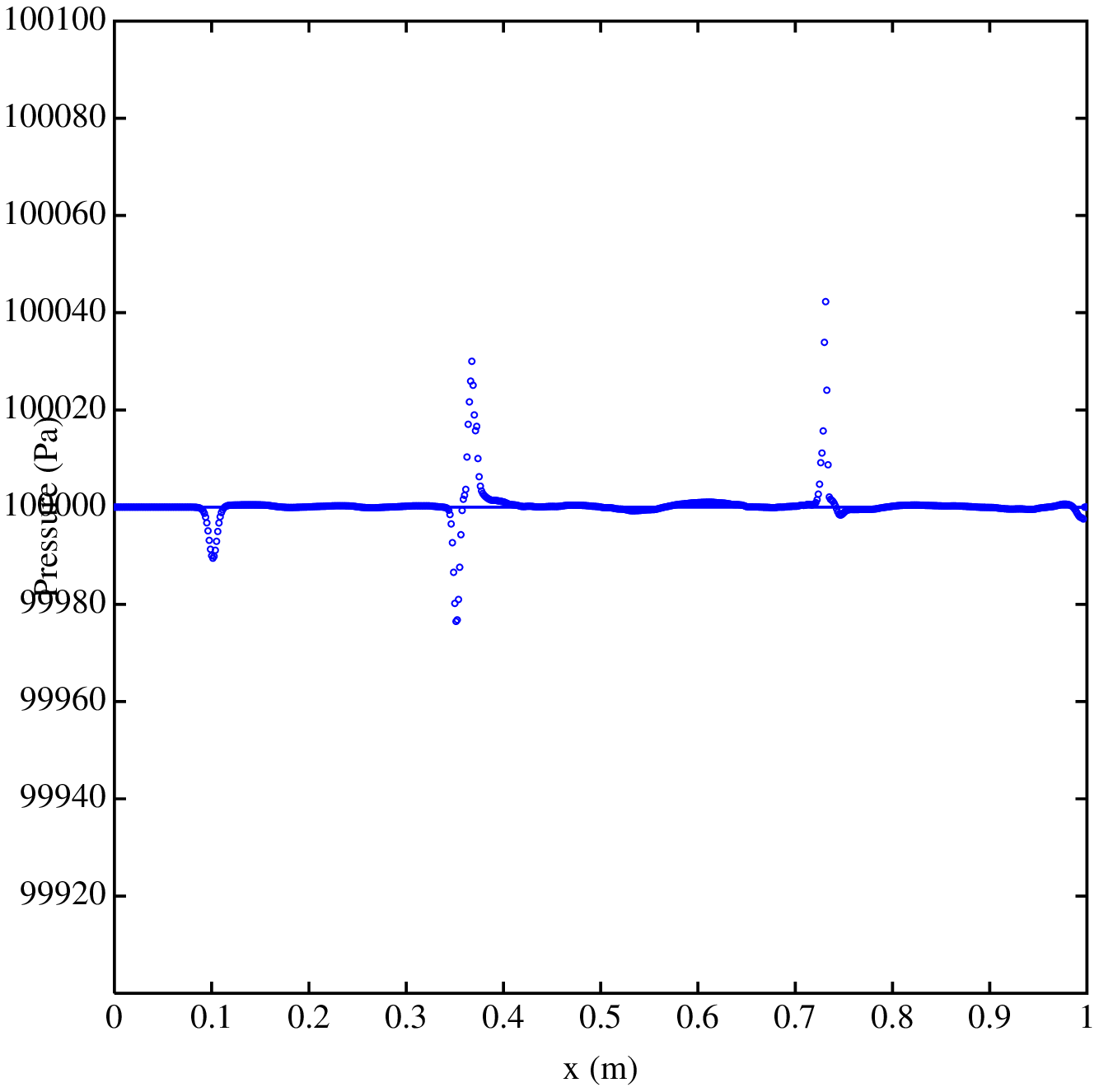}&  
\includegraphics[width=5.0cm,height=5.0cm]{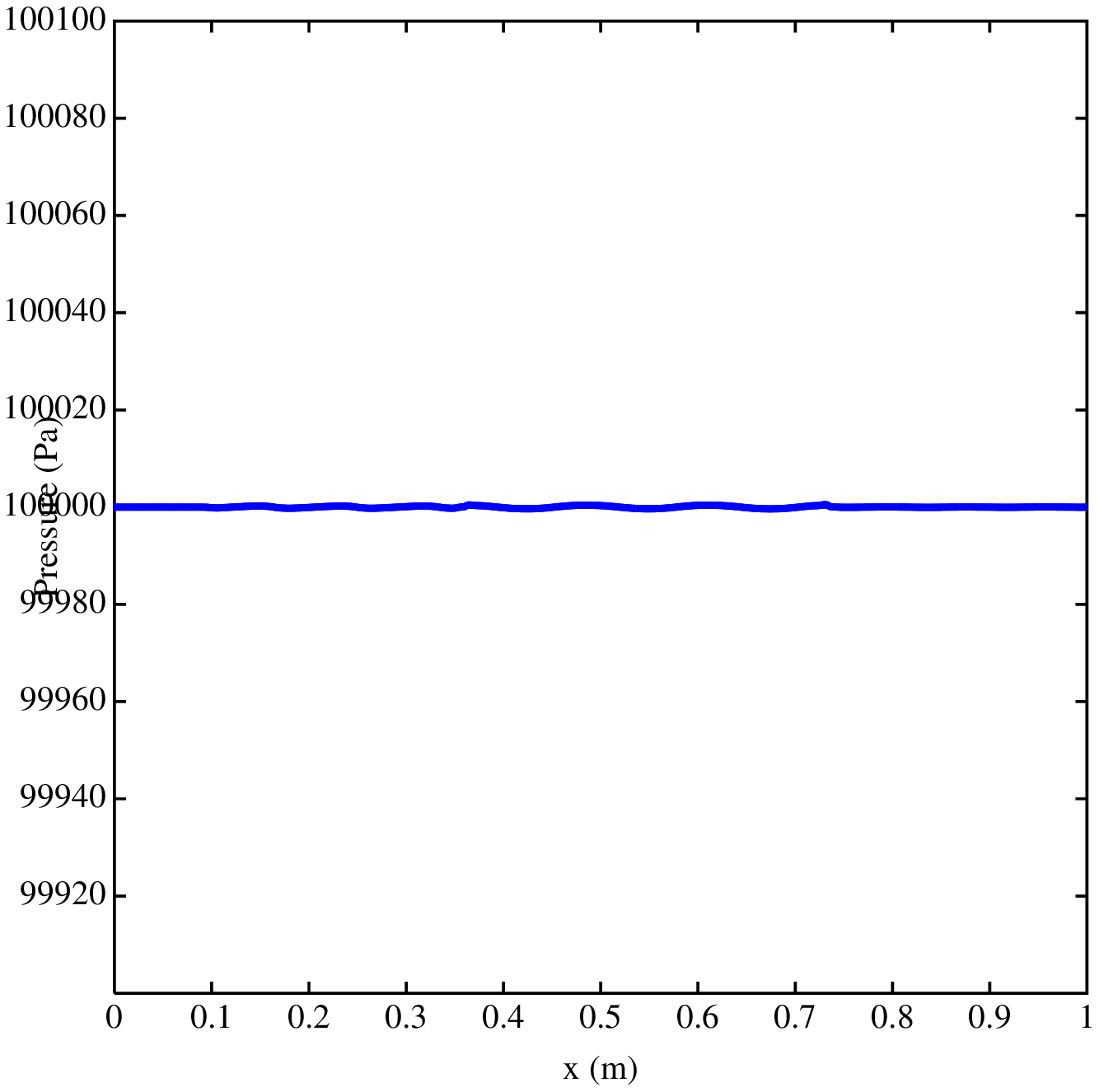}\\ 
Velocity (GFM) & Velocity (ESIM)\\ 
\includegraphics[width=5.0cm,height=5.0cm]{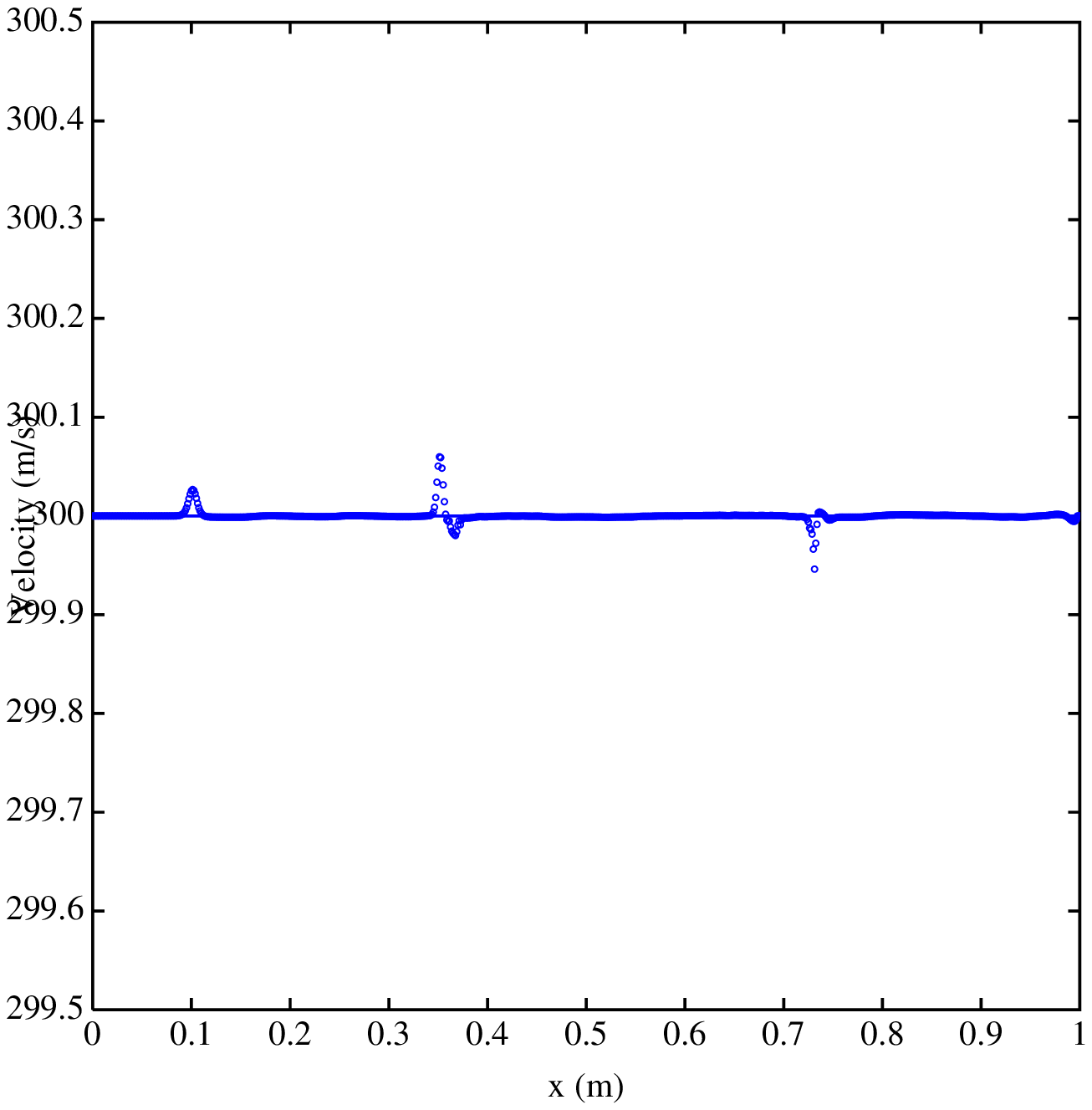}&  
\includegraphics[width=5.0cm,height=5.0cm]{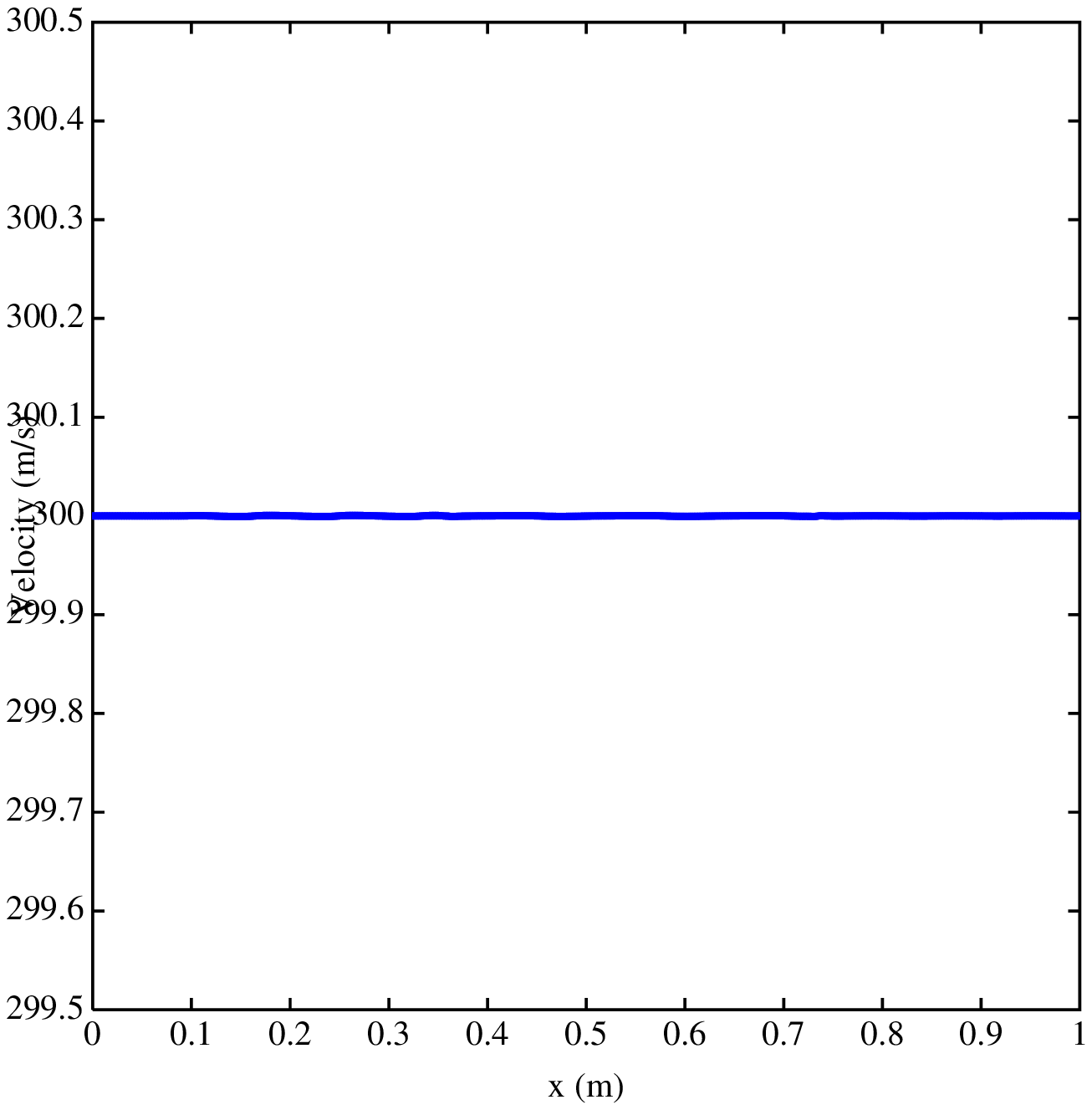} 
\end{tabular} 
\caption{Test 3-b: 800 grid points. ENO-3 coupled with the GFM (left column) and with the ESIM (right column). Exact values: solid line; numerical values: points (note that the scales for $p$ and $u$ are magnified).}                                                                                                                                                                                                                                                                                                                                                                                                                                                                                         
\label{Test3_800} 
\end{center} 
\end{figure} 
 
\begin{table}[htbp] 
\begin{center} 
\begin{tabular}{|cccc|cccc|} 
\hline 
         &         &                &                &          &         &                &               \\  
Method   &  $N_x$  &   $L_1$ error  &   $L_1$ order  &  Method  &  $N_x$  &   $L_1$ error  &   $L_1$ order \\  
         &         &                &                &          &         &                &               \\  
\hline 
\hline 
         &         &                &                &          &         &                &               \\  
         &   100   &    6.44e-3     &       -        &          &   100   &     3.76e-3    &      -        \\ 
ENO-3    &   200   &    1.62e-3     &   {\bf 1.99}   &   ENO-3  &   200   &     4.74e-4    &  {\bf 2.98}   \\ 
+        &   400   &    6.86e-4     &   {\bf 1.24}   &     +    &   400   &     6.42e-5    &  {\bf 2.88}   \\      
GFM      &   800   &    2.40e-4     &   {\bf 1.51}   &   ESIM   &   800   &     8.62e-6    &  {\bf 2.89}   \\      
         &  1600   &    6.44e-5     &   {\bf 1.89}   &          &  1600   &     9.50e-7    &  {\bf 3.18}   \\     
         &  3200   &    2.05e-5     &   {\bf 1.65}   &          &  3200   &     1.21e-7    &  {\bf 2.98}   \\     
         &         &                &                &          &         &                &               \\        
\hline 
\end{tabular} 
\end{center} 
\label{TabErreur} 
\hspace{0.5cm} 
\caption{Measures of convergence in Test 3.} 
\end{table} 

\begin{table}[htbp] 
\begin{center} 
\begin{tabular}{|c|c|cc|cc|cc|} 
\hline 
          &        &                  &           &                      &            &                    &            \\ 
Method    & $N_x$  &  $\Delta(\rho)$  &   Order   &   $\Delta(\rho\,u)$  &    Order   & $\Delta(\rho\,e)$  &  Order     \\ 
          &        &                  &           &                      &            &                    &            \\ 
\hline 
\hline 
          &        &                  &            &                     &            &                    &            \\ 
          &  100   &    7.25e-1       &    -       &        2.21e+2      &    -       &      3.22e+4       &    -       \\ 
ENO-3     &  200   &    3.82e-1       & {\bf 0.92} &        1.14e+2      & {\bf 0.95} &      1.72e+4       &  {\bf 0.90}\\  
 +        &  400   &    2.21e-1       & {\bf 0.79} &        6.65e+1      & {\bf 0.77} &      9.97e+3       &  {\bf 0.78}\\ 
GFM       &  800   &    1.27e-1       & {\bf 0.80} &        3.83e+1      & {\bf 0.79} &      5.74e+3       &  {\bf 0.79}\\  
          & 1600   &    7.10e-2       & {\bf 0.84} &        2.13e+1      & {\bf 0.84} &      3.19e+3       &  {\bf 0.84}\\ 
          & 3200   &    3.61e-2       & {\bf 0.97} &        1.08e+1      & {\bf 0.98} &      1.62e+3       &  {\bf 0.97}\\ 
          &        &                  &            &                     &            &                    &            \\                 
\hline 
          &        &                  &            &                     &            &                    &            \\ 
          &  100   &    1.98e-1       &    -       &        5.96e+1      &    -       &      8.94e+3       &    -       \\ 
ENO-3     &  200   &    4.69e-2       & {\bf 2.08} &        1.40e+1      & {\bf 2.09} &      2.11e+3       &  {\bf 2.08}\\  
 +        &  400   &    1.07e-2       & {\bf 2.13} &        3.23e+0      & {\bf 2.11} &      4.85e+2       &  {\bf 2.12}\\ 
ESIM      &  800   &    2.75e-3       & {\bf 1.96} &        8.27e-1      & {\bf 1.96} &      1.24e+2       &  {\bf 1.96}\\  
          & 1600   &    6.82e-4       & {\bf 2.01} &        2.37e-1      & {\bf 1.80} &      3.41e+1       &  {\bf 1.86}\\ 
          & 3200   &    1.54e-4       & {\bf 2.14} &        6.72e-2      & {\bf 1.81} &      9.36e+0       &  {\bf 1.86}\\ 
          &        &                  &            &                     &            &                    &            \\                 
\hline 
\end{tabular} 
\end{center} 
\hspace{0.5cm} 
\label{TabCons}
\caption{Conservation errors in Test 3.} 
\end{table} 

We consider a 1 m long domain with two material interfaces initially at $\alpha_0=0.160$ m and $\alpha_1=0.526$ m. The pressure and the velocity are initially constant: $u(x,0)=300$ m/s, $p(x,0)=10^5$ Pa. The density is initially 
$$ 
\rho(x, 0)= 
\left\{ 
\begin{array}{l} 
1+0.3 \,\sin(50\,(x-\alpha_0)) \mbox{ kg/m}^3\mbox{ if } \alpha_0 \leq x < \alpha_1,\\ 
\\ 
1 \mbox{ kg/m}^3\mbox{ else}. 
\end{array} 
\right. 
$$ 
Physical parameters are 
$$(\gamma,p_{\infty})= 
\left\{ 
\begin{array}{l} 
\gamma_0= 1.40,\,p_{\infty\,0}=10^4 \mbox{ Pa}\quad \mbox{ if} \quad x \leq \alpha_0, \\ 
\\ 
\gamma_1= 1.67,\,p_{\infty\,1}=10^5 \mbox{ Pa}\quad \mbox{ if} \quad \alpha_0<x \leq \alpha_1, \\ 
\\ 
\gamma_2= 1.40,\,p_{\infty\,2}=10^4 \mbox{ Pa}\quad \mbox{ if} \quad x > \alpha_1. 
\end{array} 
\right. 
$$
This configuration amounts to an advection equation for $\rho$; material interfaces are advected at the velocity $u$; $p$ and $u$ remain theoretically constant. Because of the discontinuous physical parameters, the entropy $S$ is discontinuous at $\alpha_0$ and at $\alpha_1$. Numerical experiments are performed with ENO-3, coupled with the GFM or with the ESIM. 
  
Figures \ref{Test3_200} and \ref{Test3_800} show exact values and numerical values of $\rho$, $S$, $p$, and $u$ at $t=6.62\,10^{-4}$ s, respectively for $N_x=200$ grid points (hence 25 grid points by wavelength) and $N_x=800$ grid points (hence 100 grid points by wavelength). Notice that the zeroth-order extrapolation of $S$ used by the GFM result in jumps
of $\rho$ at material interfaces, that are advected with the flow; these glitches are also transferred to $u$, and $p$, near $x=0.35$ m and $x=0.75$ m, and act as sources of acoustic noise. These glitches still exist with 800 grid points (figure \ref{Test3_800}) with the GFM. On the other hand, no entropy or density glitches are observed when the ESIM is used
instead. 

We must mention that ENO reconstructions of a sinusoidal profile might produce oscillations on the level of the truncation
error. These small spurious oscillations are seen in $p$ and $u$, which should have flat profiles, and can be observed even without material interfaces. When these oscillations interact with the GFM-produced glitches at material interfaces, they are amplified and lead also to spurious acoustic waves. These oscillations are entirely reconstruction dependent
and, since they are of the order  of the truncation error (see also our convergence measures), they are essentially not visible on 800 grid points when the ESIM is used (figure \ref{Test3_800}).

Measures of convergence are provided in table 5.1. Coupled with the ESIM, the ENO-3 scheme maintains third-order convergence. Coupled with the GFM, the ENO-3 scheme looses accuracy and shows a 1.6 order of convergence. Table 5.2 shows measures of conservation errors induced by the GFM and by the ESIM, for various values of $N_x$ (see subsection \ref{SEC_REMARK}). To do so, we compute 
\begin{equation} 
\Delta\,\boldsymbol{U} (T, N_x)= 
\max_{n=0,...,N_t}  
\left | 
\sum_{i=i_0}^{i_1}  
\left( 
\boldsymbol{U}_i^n -\boldsymbol{U}_i^0 
\right) 
+n\frac{\textstyle \Delta\,t}{\textstyle \Delta\,x}  
\left(\boldsymbol{f}\left(\boldsymbol{U}_{i_1}^n \right)- 
\boldsymbol{f}\left(\boldsymbol{U}_{i_0}^n \right)\right) 
\right | 
\label{MesCons} 
\end{equation} 
at $T=1.05\,10^{-3}$ s, where $\boldsymbol{f}$ is the flux function (\ref{FLUX_EULER}) 
, and $N_t=\mbox{ Trunc }(T / \Delta\,t)$. The domain of measure is bounded by $i_0=10$ and $i_1=N_x-10$. Getting fiable and meaningfull measures requires some care. Indeed, errors of conservativity vary a lot with the position of $\alpha$ inside the meshing, hence with $t$. The "max" in (\ref{MesCons}) ensures almost-steady values of $\Delta\,\boldsymbol{U}(T,N_x)$ and reliable measures.  
 
A fully conservative scheme would satisfy $\Delta\,\boldsymbol{U}(T,N_x)=0$ for all values of $T$ and $N_x$ (see e.g. (12.33) in \cite{LEV90}). Here, errors are non-null, but they decrease with $\Delta\,x$. Conservation errors induced by the ESIM are much smaller than that induced by the GFM. The table 5.2 indicates first-order conservation errors for the GFM, and second-order conservation errors for the ESIM. 

 Measures of convergence and conservativity have been done also with the WENO-5 scheme coupled with the GFM or with the ESIM. The results are essentially the same than in tables 5.1 and 5.2. To get a fifth-order accurate scheme (as in \cite{PIRAUX1}) and a higher order of conservation would require to develop a higher order ESIM, which is not investigated in the present study. 

\subsection{Test 4: nonlinear acoustics} 
 
\begin{figure}[htbp] 
\begin{center} 
\includegraphics[width=5.5cm,height=5.5cm]{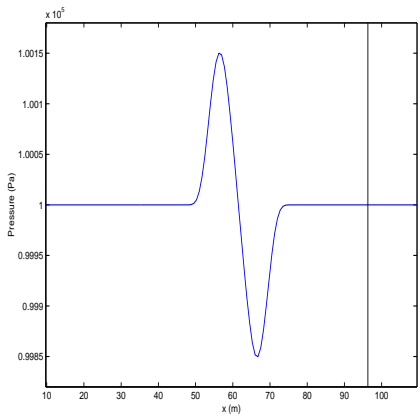}\\ 
\caption{Initial values of the pressure $p$ at $t_0=5.1\,10^{-2}$ s in Test 4.} 
\label{Test4_Init} 
\end{center} 
\end{figure} 
\begin{figure}[htbp] 
\begin{center} 
\begin{tabular}{cc} 
(a) & (b) \\ 
\includegraphics[width=5.5cm,height=5.5cm]{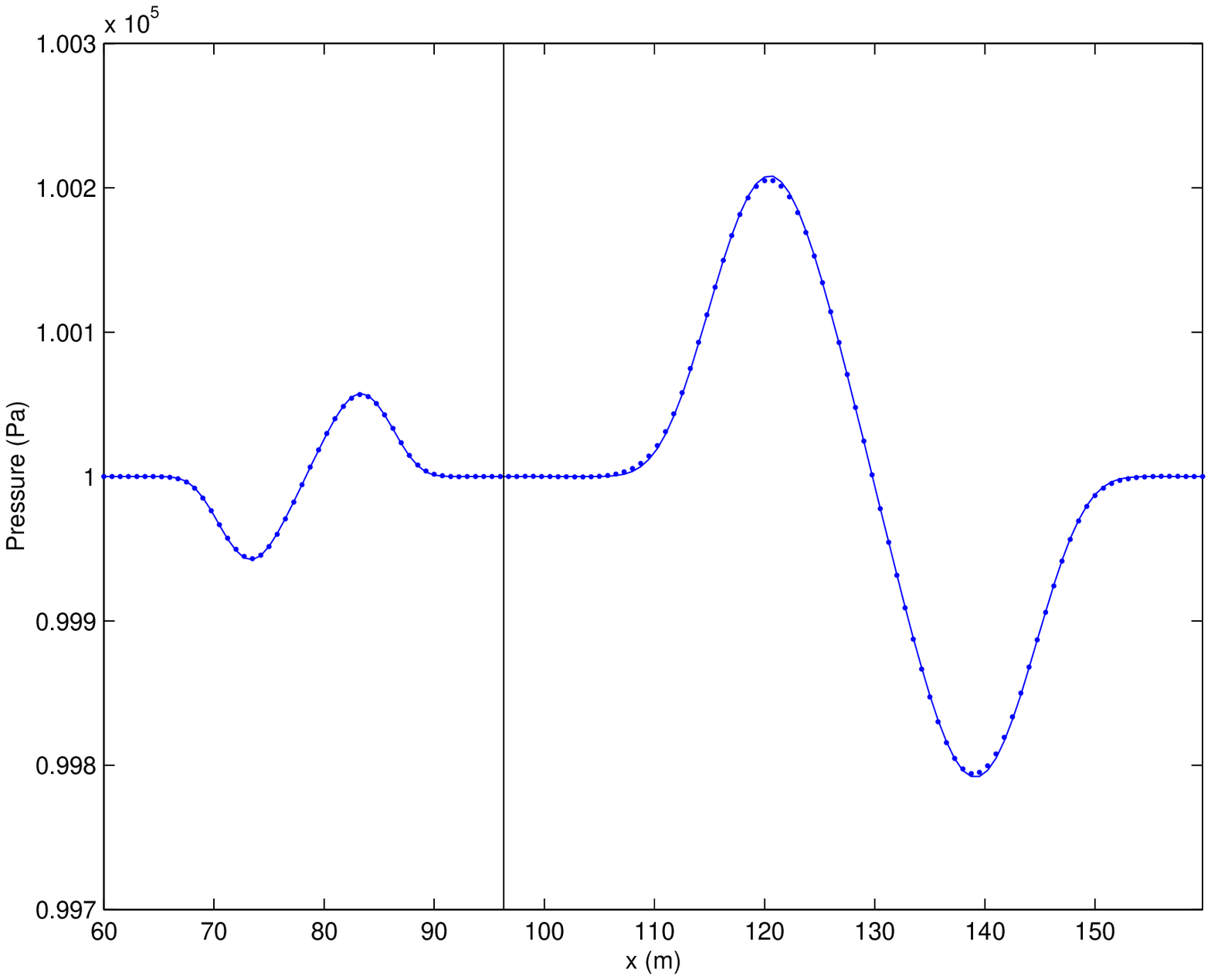}&  
\includegraphics[width=5.5cm,height=5.5cm]{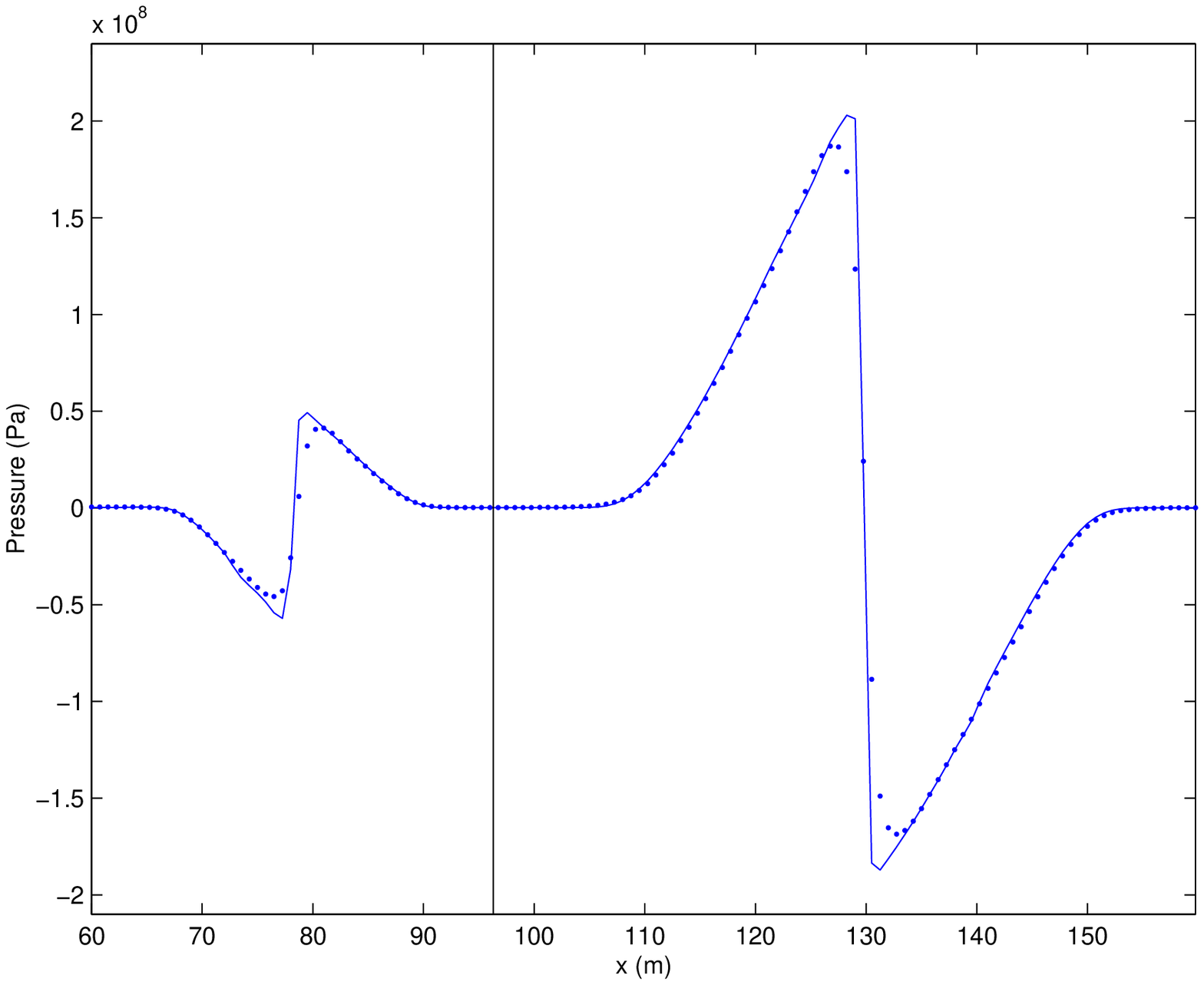}\\ 
(c) & (d) \\ 
\includegraphics[width=5.5cm,height=5.5cm]{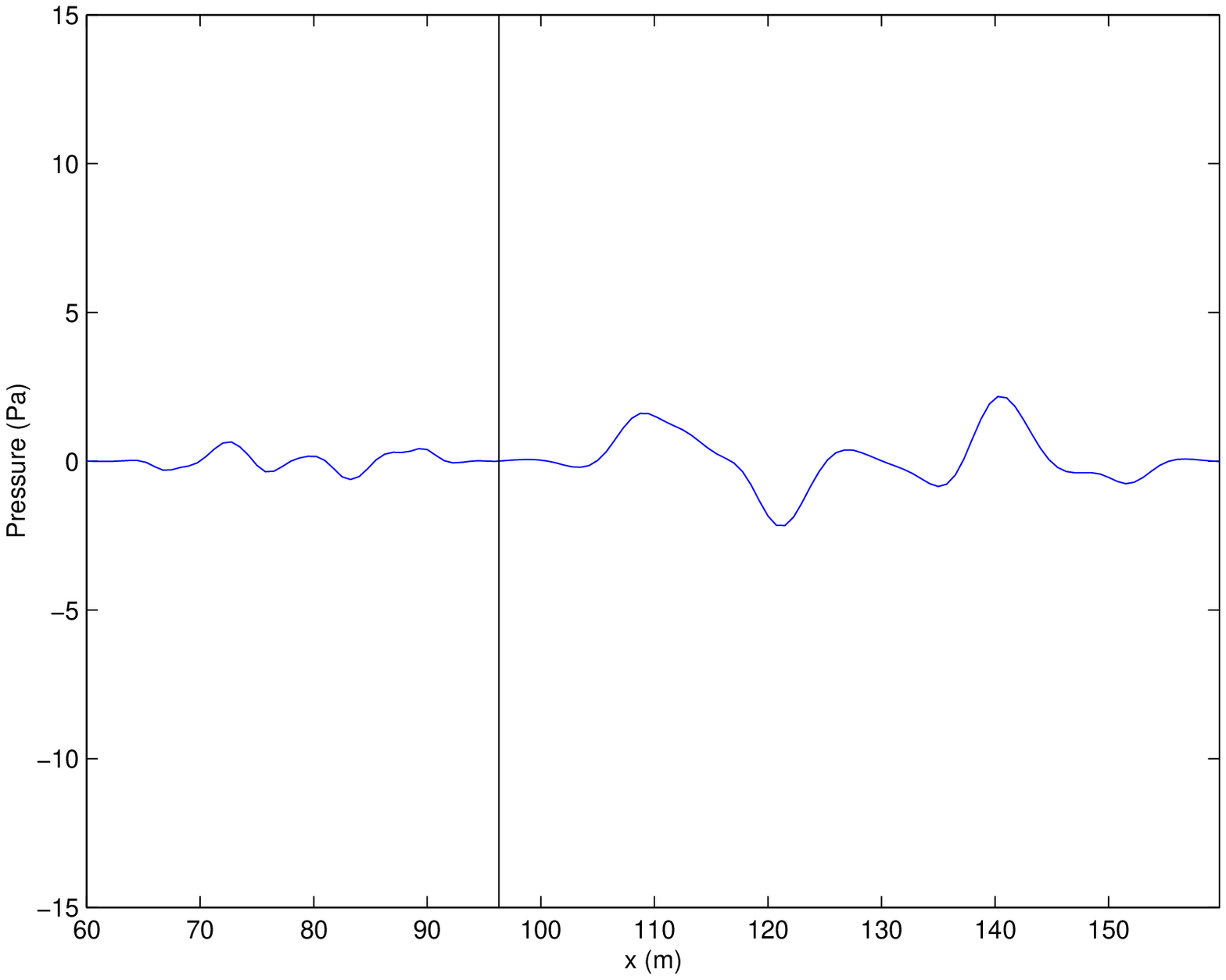}&  
\includegraphics[width=5.5cm,height=5.5cm]{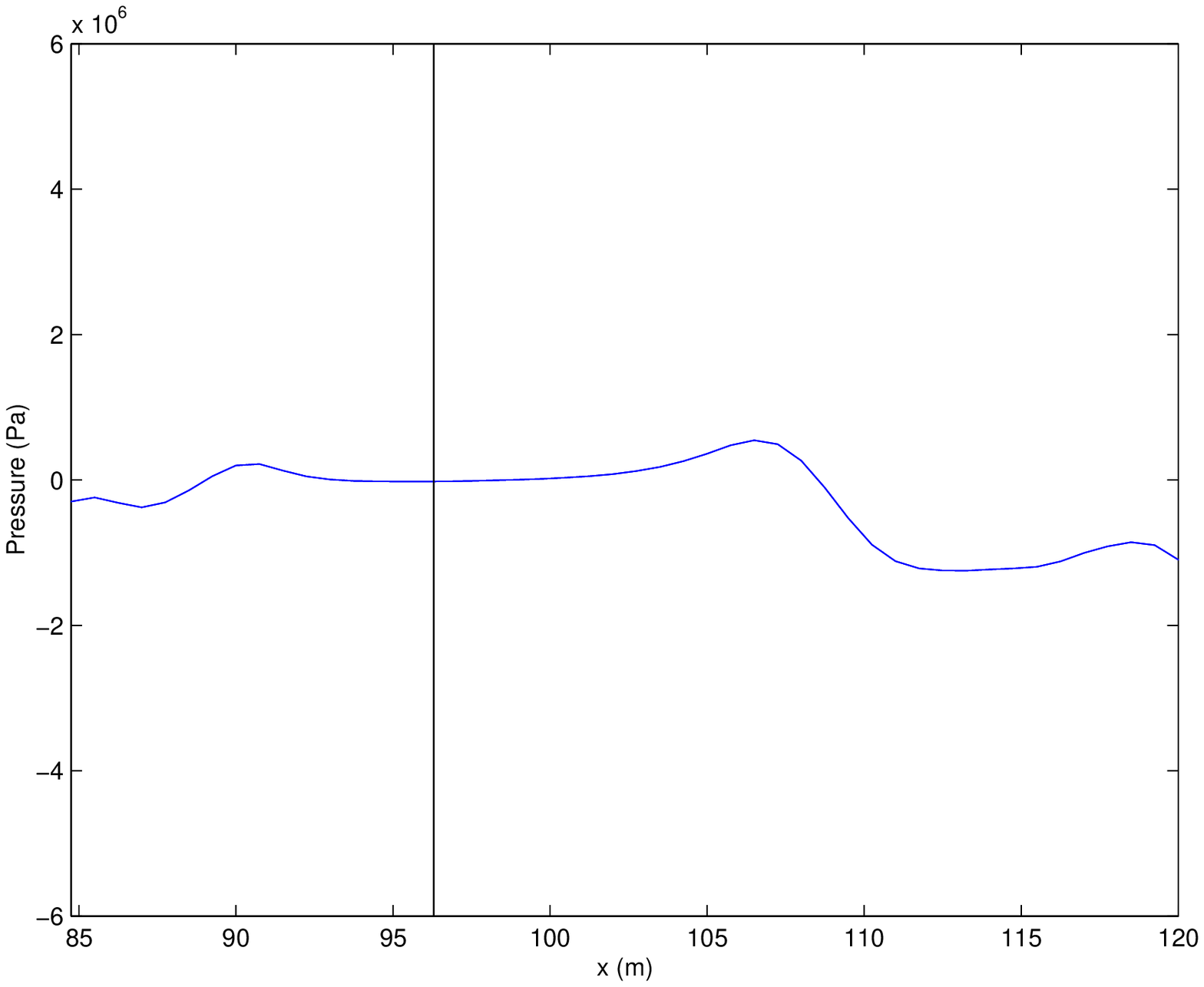}\\ 
(e) & (f) \\ 
\includegraphics[width=5.5cm,height=5.5cm]{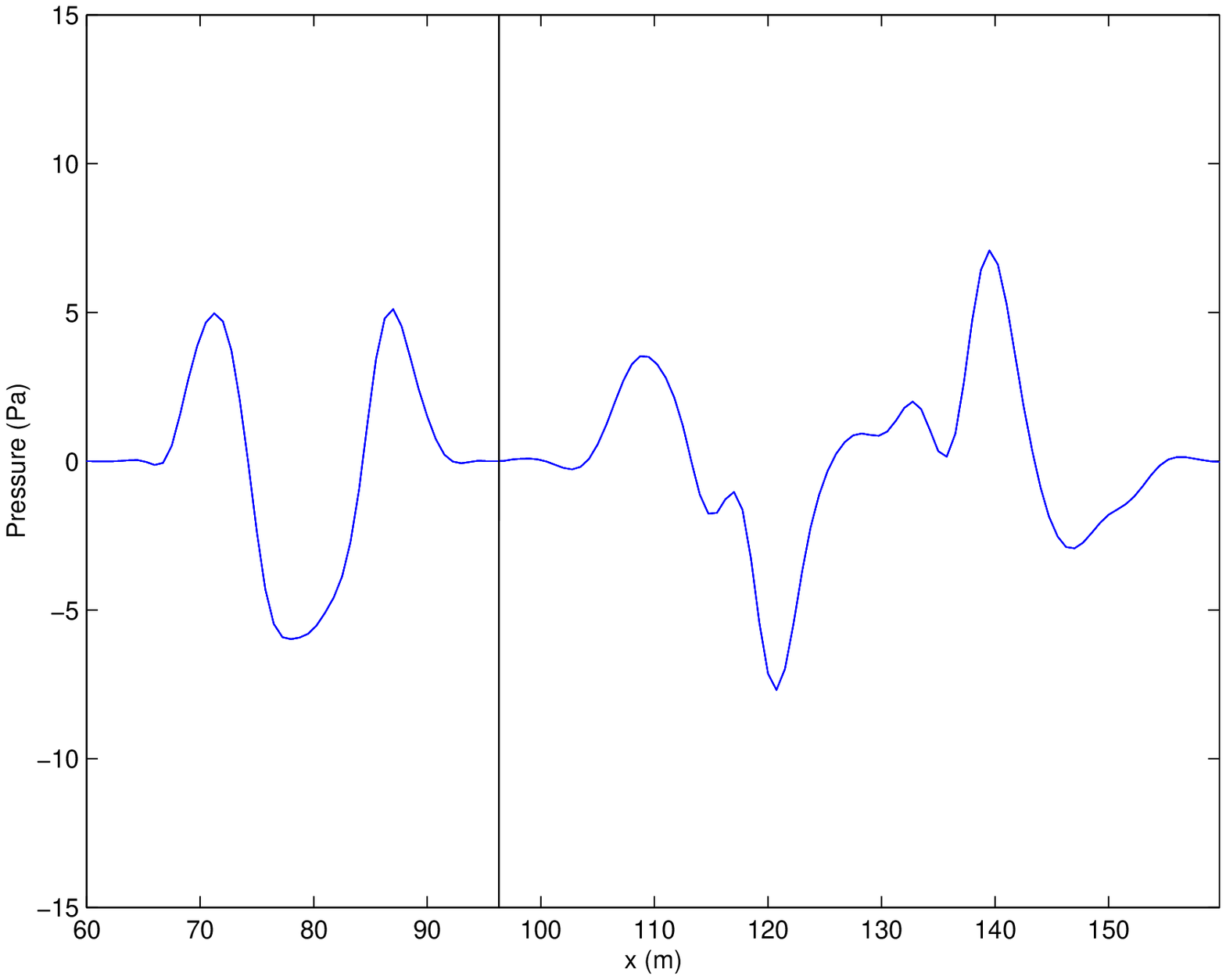}&  
\includegraphics[width=5.5cm,height=5.5cm]{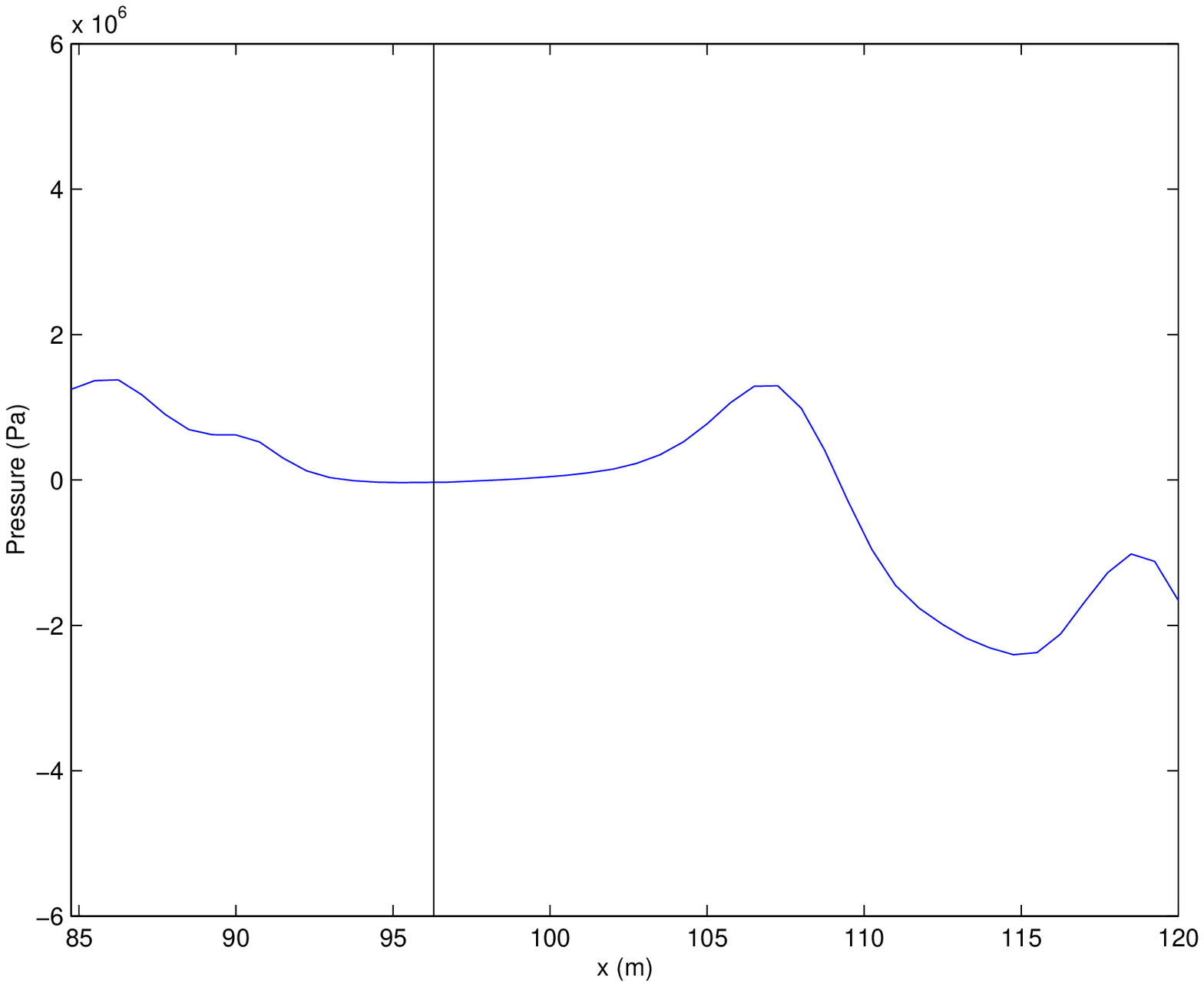} 
\end{tabular} 
\caption{Test 4. $\varepsilon = 10^{-3}$ (left column), $\varepsilon = 1$ (right column) at $t_1=8.61\,10^{-2}$ s. Snapshots of fine grid solution (solid line) and numerical values (dotted line) of $p$ with the ESIM: (a), (b). Errors with the ESIM: (c), (d). Errors with the GFM: (e), (f).} 
\label{Test4} 
\end{center} 
\end{figure} 
As a last example, we consider a rich-structured wave interacting with a stationary material interface. This example is a generalization of the Test 1 studied in \cite{PIRAUX1}: for small amplitude initial data, the simple linear equations of acoustics used in \cite{PIRAUX1} are valid, and the wave remains smooth. Higher amplitudes of initial data lead to a nonlinear problem, where shocks can develop. 
 
To illustrate both cases, we use exactly the same physical parameters than in \cite{PIRAUX1}. On a 300 m long domain, a material interface separates piecewise constant values, initially motionless 
$$ 
\left\{ 
\begin{array}{l} 
\rho_0= 1000\mbox{ kg/m}^3,\, p_0=10^5\mbox{ Pa },\, u_0=0.0\mbox{ m/s }, \gamma_0=3,\, p_{\infty\,0}=7.50\,10^8\mbox{ Pa},\\ 
\\ 
\rho_1= 1200\mbox{ kg/m}^3,\, p_1=10^5\mbox{ Pa },\, u_1=0.0\mbox{ m/s }, \gamma_1=4,\, p_{\infty\,1}=2.35\,10^9\mbox{ Pa}. 
\end{array} 
\right. 
$$ 
From (\ref{SSpeed}), one deduces sound speeds: $c_0$ = 1500 m/s, and $c_1$ = 2800 m/s. These physical parameters correspond respectively to water and Plexiglass under atmospheric pressure. A perturbation $\Delta\,\boldsymbol{W}_0(x)$ is added to initial data on medium $\Omega_0$, which is an exact solution of acoustics and leads to a right-going wave
\begin{equation} 
\Delta\,\boldsymbol{W}_0(x)=\varepsilon\,f\left(t_0-\frac{\textstyle x}{\textstyle c_0}\right) 
\,^T\left( 
-\frac{\textstyle p_0}{\textstyle c_0^2},-\frac{\textstyle p_0}{\textstyle \rho_0\,c_0},-p_0 
\right). 
\label{DU0} 
\end{equation} 
The function $f$ is a $C^5$ spatially-bounded sinusoid 
\begin{equation} 
f(\xi)= 
\left\{ 
\begin{array}{l} 
\displaystyle \sum_{k=1}^{q} a_k\,\sin(\beta_k\,\omega_c\,\xi) 
\quad \mbox{ if  }\, 0<\xi<\frac{\textstyle 1}{\textstyle f_c},\\ 
\\ 
0 \,\mbox{ else},  
\end{array}  
\right.                        
\label{SIGNAL} 
\end{equation} 
with $\beta_k=2^{k-1}$, $\omega_c=2\pi\,f_c$; the coefficients $a_k$ are: $a_1=1$, $a_2=-21/32$, $a_3=63/768$, $a_4=-1/512$. The central frequency is $f_c=50$ Hz. Elementary calculations show that  
\begin{equation} 
\frac{\textstyle \Delta\,\rho_0}{\textstyle \rho_0}\sim  
\frac{\textstyle \Delta\,u_0}{\textstyle c_0}\sim 
\frac{\textstyle \Delta\,p_0}{\textstyle p_0}\sim   
\varepsilon. 
\end{equation} 
For $\varepsilon=0$, the solution remains a stationary material interface. For small values of $\varepsilon$ (typically $\varepsilon = 10^{-3}$), the acoustics limit is valid and one obtains a pure right-going wave. For higher values of $\varepsilon$ (typically $\varepsilon = 0.1$), the perturbation (\ref{DU0}) does not satisfy exactly the nonlinear Euler equations: the wave separates into a weak left-going wave and a right-going wave. In both cases, the right-going wave is reflected and transmitted by the material interface. The analytical solution is not detailed in the linear case \cite{THESE_MOI}. No analytical solution is available in the nonlinear case: one computes an "exact solution" on a fine grid of 3200 grid points. Numerical experiments are performed with $N_x$ = 400 grid points, which is 40 points in the wavelength $\lambda_c=c_0/f_c$, and $t_{0}=5.1\,10^{-2}$ s. Initial values of the $p$ at $t_0$ are shown in figure \ref{Test4_Init}, for $\varepsilon = 10^{-3}$. 
 
Figure \ref{Test4} shows results at $t_1=8.61\,10^{-2}$ s, after 200 time steps. Left and right columns concern respectively $\varepsilon = 10^{-3}$ and $\varepsilon = 0.1$. Figures \ref{Test4} (a) and (b) show numerical values and exact values of $p$ computed by WENO-5 coupled to the ESIM. Logically, (a) is similar to figure 3-(d) of \cite{PIRAUX1}. In the nonlinear case (b), the reflected and transmitted waves have developed a shock. In both cases, the agreement between exact values and numerical values is excellent. 
 
To see clearly differences between the ESIM and GFM treatments, we display errors for both methods: with the ESIM in (c) and (d), with the GFM in (e) and (f). In the nonlinear cases (d) and (f), errors are displayed from $x$ = 85 m to $x$ = 120 m, to avoid the shock area. In both cases, the graphs confirm that the ESIM is more accurate than the GFM. 
 
Unlike in \cite{PIRAUX1}, the material interface is allowed to move with the flow. For $\varepsilon = 10^{-3}$, the measured movement is lower than $10^{-7}$ m: so, the approximation of stationary material interfaces is justified. For $\varepsilon = 0.1$, the material interface moves from 96.3 m to 96.114 m, and then it comes back to the initial position 96.3 m: it is logical, since the velocity is symetric with respect to zero. 
 
\section{Conclusion}
 
We have proposed an extension of the "Explicit Simplified Interface Method" (ESIM), previously developed in acoustics \cite{LOMBARD1,LOMBARD2,PIRAUX1}, to treat material interfaces in 1D multicomponent Euler flows. 

The method enforces the numerical solution to satisfy zero-order and first-order jump conditions at the material interface and can be coupled with the user's favourite high-order shock-capturing scheme for single component flows. It behaves as robustly as the GFM \cite{GFM} in numerical simulations involving flat states, while it displays a superior performance in flows with rich structures.

Our numerical simulations show that the interface treatment we propose guarantees a numerical solution with no unphysical numerical artifacts due to the material interface. The behavior observed for the numerical errors is similar to that observed for the same underlying method when applied to a flow without material interfaces. 

This paper was focused on 1D flows: applying the ESIM to 2D cases is a challenging project, subject of future works. Some key tools for the 2D implementation have been validated already in the simple case of 2D linear acoustics with stationary interfaces \cite{LOMBARD2}. Other ingredients (such as multidimensional level-set-based extrapolations) may be found in \cite{ASLAM,IIM_HOU}.

\vspace{0.5cm} 
{\bf Acknowledgments.} 
The first author thanks the European network HYKE, contract HPRN-CT-2002-00282, to have financed his postdoctoral position at the University of Valencia, where the present research was done. The second author also acknowledges partial support from the spanish MCYT-BFM-2001-2814. The authors would like to thank the anonymous referees for their comments and for pointing to us several useful references.

\end{document}